\documentclass[reprint,amsmath,amssymb,aps]{revtex4-2}
\usepackage{amsfonts}
\usepackage{graphicx}
\usepackage{bm}
\usepackage{bbold}
\usepackage{hyperref}
\usepackage{color}
\usepackage{tabularx}
\usepackage{url}

\begin{document}

\title{Direct demonstration of bulk-boundary correspondence in higher-order topological superconductors with chiral symmetry}
\author{Xiaoyu Zhu}
\affiliation{International Center for Quantum Design of Functional Materials (ICQD), Hefei National Research Center for Physical Sciences at the Microscale, University of Science and Technology of China, Hefei, Anhui 230026, China}
\affiliation{School of Physics, MOE Key Laboratory for Non-equilibrium Synthesis and Modulation of Condensed Matter, Xi'an Jiaotong University, Xi'an, Shaanxi 710049, China}

\date{\today}

\begin{abstract}

    A higher-order topological superconductor can experience topological phase transitions driven by variations in a bulk parameter without closing the bulk gap. This presents a challenge in establishing a direct bulk-boundary correspondence, as conventional bulk invariants change only upon the closure of the bulk gap. Our study of two-dimensional higher-order phases in the DIII and BDI symmetry classes, both characterized by chiral symmetry, demonstrates that zero-energy crossings facilitate a direct connection between the bulk Hamiltonian and Majorana zero modes at corners. These crossings, emerging as boundary conditions vary, can be identified from the bulk Hamiltonian. For both classes, we introduce a pair of topological invariants derived from these zero-energy crossings to characterize the higher-order topology. Phases in which at least one invariant assumes a nonzero value are anticipated to host Majorana corner modes. Moreover, these invariants may change with the closure of either bulk or edge gaps, thereby providing a clear and direct demonstration of bulk-boundary correspondence in higher-order phases. Our findings offer a promising framework for systematically exploring higher-order topology through boundary condition modulation.

\end{abstract}

\maketitle

\section{Introduction}

Topology in superconducting systems is typically characterized by invariants derived from bulk states under the assumption of periodic boundary conditions (PBC) \cite{qi2011,chiu2016}. In most scenarios, a nontrivial topological invariant guarantees the emergence of localized Majorana modes \cite{read2000,kitaev2001,wilczek2009,alicea2012,stanescu2013,beenakker2013,elliott2015,Aguado2017} at the boundaries (edges or surfaces), which are introduced by ``cutting" the periodic system. This phenomenon, known as bulk-boundary correspondence, is a defining feature of topological states of matter. The recognition of higher-order topology \cite{benalcazar2017a,benalcazar2017,langbehn2017,song2017,schindler2018,schindler2018a,benalcazar2022} extends this correspondence, leading to the expectation that Majorana modes in superconductors may also appear at intersections (corners or hinges) of adjacent boundaries. This implies that a nontrivial higher-order topological superconductor \cite{khalaf2018,zhu2018,yan2018,wang2018b,wang2018a,yan2019,zhang2019a,ghosh2021,scammell2022} manifests Majorana modes when cut at least twice along different directions. While this higher-order extension broadens the potential material base within the topological family, it also presents challenges in fully understanding the connection between bulk properties and boundary signatures.

A pertinent question is whether one may identify topologically protected Majorana modes directly from the bulk Hamiltonian in higher-order topological superconductors. The answer to this question becomes nuanced when considering the influence of crystalline symmetries. It is well known that crystalline symmetry, such as mirror symmetry, can protect higher-order topology unless the bulk gap closes \cite{geier2018}. States exhibiting this characteristic are termed intrinsic higher-order topological states \cite{trifunovic2019}, resembling topological crystalline states, wherein characterizing nontrivial topology from the bulk states of a periodic system, using symmetry indicators for instance \cite{skurativska2020,ono2020,takahashi2020,hsu2020,kruthoff2017,tang2019,zhang2019,vergniory2019,zhang2022a,roberts2020,huang2021,tang2022a}, is plausible. However, higher-order states also distinguish themselves from topological crystalline states in that their boundary signatures can persist even under perturbations breaking related crystalline symmetries \cite{zhu2019,khalaf2021,tiwari2020,volpez2019,wu2020}. In this circumstance, variations in a bulk parameter may drive the system across topological phase transitions without closing the bulk gap, making it challenging to establish a direct bulk-boundary correspondence.

In our previous work \cite{wang2023c} on two-dimensional (2D) superconductors in the D symmetry class, we discovered that zero-energy crossings \cite{beenakker2013b,kimme2016} in the Bogoliubov-de Gennes (BdG) energy spectrum serve as robust indicators for higher-order topology. These crossings manifest as the 2D system transitions from toroidal to cylindrical geometry. Topological phase transitions, occurring where either the bulk or edge gap closes, are accompanied by the emergence or disappearance of zero-energy crossings. Importantly, these crossings could be readily identified from the bulk Hamiltonian, thereby acting as a bridge connecting the bulk Hamiltonian and Majorana corner modes. In this study, we focus on the DIII and BDI symmetry classes, and examine the higher-order bulk-boundary correspondence within the same framework. In contrast to the D class, the topological charges associated with these crossings --- well defined due to the presence of chiral symmetry in the two classes --- also play a pivotal role in determining the higher-order topological invariants.

The rest of this article is organized as follows: In Sec. \ref{sec:II}, we set up the general framework and introduce the boundary-modulated Hamiltonian that is central to our study. In Sec. \ref{sec:III}, we demonstrate how a pair of higher-order topological invariants, which are derived from zero-energy crossings, characterize the higher-order topology in the DIII class, and then take a specific model as an example for illustration. In Sec. \ref{sec:IV} we investigate the higher-order bulk-boundary correspondence in the BDI class following the same procedure as in the DIII class. Finally, in Sec. \ref{sec:V} we provide additional discussions and a brief summary to our work.

\begin{figure}[t]
    \includegraphics[width=0.48\textwidth]{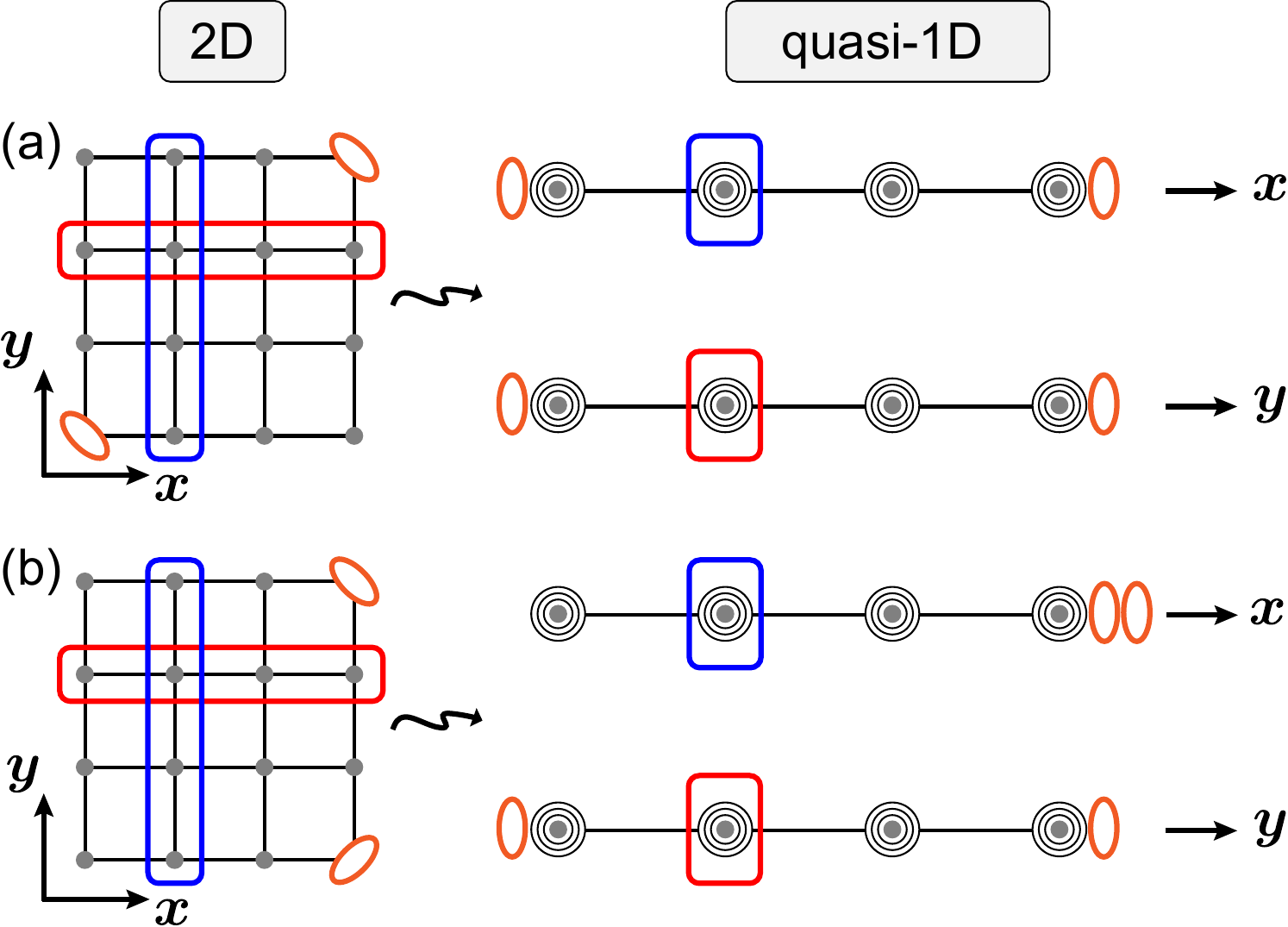}
    \caption{Illustration of the equivalence between higher-order topology in 2D systems and first-order topology in quasi-1D systems. (a) Majorana zero modes, indicated by the orange ovals, reside at two opposite corners in the 2D system. In this configuration, the quasi-1D systems along the $x$ and $y$ direction both exhibit nontrivial first-order topology. (b) Majorana zero modes appear at two adjacent corners. In this scenario, only the quasi-1D system oriented along the $y$ direction demonstrates nontrivial first-order topology. Lattice sites enclosed by blue (red) rectangles correspond to a single unit cell in the quasi-1D system extending along the $x(y)$ direction.}\label{fig1}
\end{figure}

\section{Boundary-modulated Hamiltonian}\label{sec:II}

We consider a generic 2D superconducting system with boundary conditions modulated in the $x$ or $y$ direction \cite{takahashi2020,tanaka2020}. The Hamiltonian is composed of two parts, given by
\begin{equation}
    \mathcal H_a(\lambda_a)  =  \mathcal H - (1-\lambda_a)\mathcal B_a, \label{eq:ham}
\end{equation}
where $a=x (y)$ and $\bar a = y(x)$. The first part, $\mathcal H$, represents the bulk Hamiltonian, describing a system with PBC in both directions. The second part, $\mathcal B_a$, introduces boundary terminations and is referred to as the \emph{boundary Hamiltonian}. $\mathcal B_a$ includes all inter-cell terms crossing the terminations (edges) that extend along the $\bar a$ direction. It is important to note that this definition of the boundary Hamiltonian does not involve intra-cell terms at boundary sites. Take the 2D spinless $p$-wave superconductor as an example. The inter-cell terms consist of nearest-neighbor hopping and pairing terms, whose amplitudes are represented by $t$ and $\Delta$ respectively. Denote $c_{i,j}$ as the annihilation operator at site labeled by $(i,j)$, and we have $\mathcal B_x = \sum_{j=1}^{N_y} t c^\dagger_{N_x,j}c_{1,j}+ \Delta c^\dagger_{N_x,j}c^\dagger_{1,j} + \text{H.c.}$, where $N_a$ refers to the number of unit cells along the $a$ direction. In general, when considering the hopping with range $r$, we need to include all the corresponding terms that cross the edge in the $y$ direction, like $\sum_{n=1}^r c^\dagger_{N_x-n+1,j}c_{r-n+1,j}+\text{H.c.}$. In this paper, we only consider boundary terminations that are commensurate with unit cells. As such, the boundary Hamiltonian is solely determined by the inter-cell terms in the bulk Hamiltonian. The real parameter $\lambda_a$ controls the boundary condition in the $a$ direction and modulates the amplitude of the boundary Hamiltonian. In particular, $\lambda_a = 1(0)$ corresponds to periodic (open) boundary condition, with $\mathcal{H}_a(\lambda_a=1(0))$ describing a toroidal (cylindrical) system. Varying $\lambda_a$ continuously in the range $[0,1]$ is akin to gradually cutting the periodic system along the $\bar a$ direction.

For the boundary-modulated Hamiltonian $\mathcal H_a$, the momentum along the $\bar a$ direction remains a good quantum number, allowing us to express its second-quantized form as
\begin{equation}
    \mathcal H_a(\lambda_a) = \sum_{k_{\bar a}} \Psi_a^\dagger(k_{\bar a}) H_a(k_{\bar a},\lambda_a)\Psi_a(k_{\bar a}),\label{eq:ham_a}
\end{equation}
where $\Psi_a(k_{\bar a})$ is the Nambu spinor, and $H_a(k_{\bar a},\lambda_a)$ is the Bloch BdG Hamiltonian, defined in the space parameterized by $(k_{\bar a},\lambda_a)$, with $k_{\bar a}\in [-\pi,\pi)$ and $\lambda_a\in[0,1]$. We may alternatively view $\mathcal{H}_a$ as a series of quasi-one-dimensional (1D) Hamiltonian with varying $\lambda_a$, which all extend along the $\bar a$ direction, as depicted in Fig. \ref{fig1}. The unit cell for this quasi-1D system has a dimension proportional to the number of lattice sites in the $a$ direction. This boundary-modulated Hamiltonian is central to our study of higher-order topology in both symmetry classes. Unless otherwise stated, we will omit the subscript of $k_{\bar a}$ and $\lambda_a$, with the understanding that the Bloch Hamiltonian $H_a(k,\lambda)$ is defined in the $(k_{\bar a},\lambda_a)$ parameter space.

To determine the higher-order topology, we begin with the cylindrical Hamiltonian $\mathcal{H}_a(\lambda=0)$, which features two edges stretching along the $\bar a$ direction. When the 2D system supports Majorana corner modes, distributed as in Fig. \ref{fig1}, the quasi-1D cylindrical system exactly resembles a nontrivial first-order topological superconductor in 1D, with stable Majorana zero modes at its ends. In this case, we may directly apply the topological invariants used for diagnosing first-order topology to the cylindrical system. When assuming nontrivial values, these invariants indicate Majorana zero modes emerging at the corners, formed by cutting the cylindrical system along the direction perpendicular to its edges. As the invariants are derived from cylindrical Hamiltonian, they fail to reveal the connection between Majorana corner modes and the 2D bulk (toroidal) Hamiltonian. To explore the bulk-boundary correspondence in higher-order phases, we shall investigate how the boundary-modulated Hamiltonian responds to the continuously varying $\lambda$.

\begin{figure}[t]
    \includegraphics[width=0.48\textwidth]{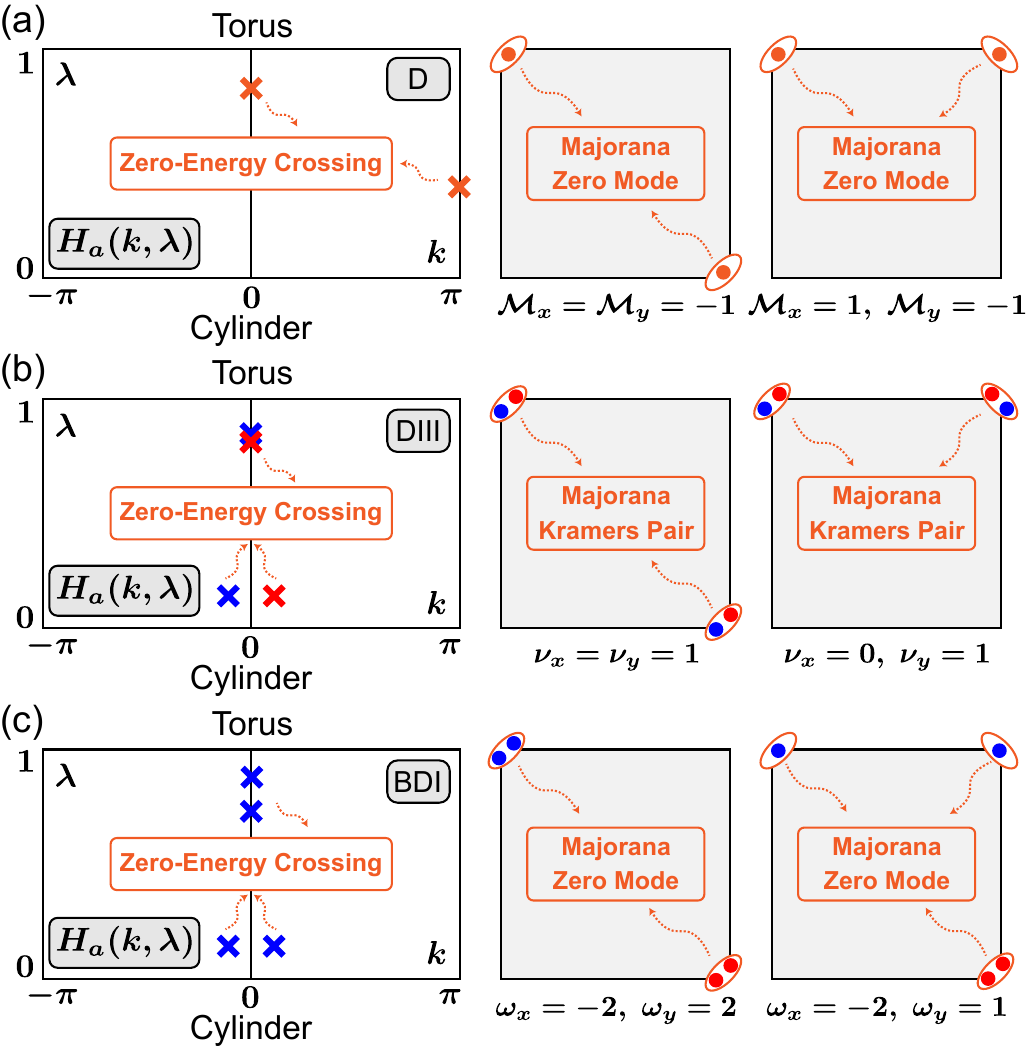}
    \caption{Zero-energy crossings in nontrivial higher-order phases of three symmetry classes. Higher-order topology in each class is characterized by a pair of topological invariants. (a) In the D class, only the crossings (marked by orange crosses) on high-symmetry lines contribute to the $\mathbb Z_2$ type topological invariants $(\mathcal M_x,\mathcal M_y)$. (b) In the DIII class, zero-energy crossings occur in pairs at $(\pm k_F,\lambda_F)$, with each pair possessing opposite topological charges, indicated by crosses in different colors. The system with one or both of the two $\mathbb Z_2$ invariants $(\nu_{x},\nu_{y})$ taking nonzero values is expected to support Majorana Kramers pairs at corners. (c) In the BDI class, zero-energy crossings off high-symmetry lines appear in pairs at $(\pm k_F,\lambda_F)$. Unlike in the DIII class, the two crossings in each pair may not have opposite topological charges. The higher-order topological invariants $(\omega_x,\omega_y)$ in this class are of $\mathbb Z$ type. In panels (b) and (c), the colors of the dots indicate the chiralities ($\pm 1$) of Majorana zero modes.}\label{fig2}
\end{figure}

In our previous study of D-class superconductors, we established that zero-energy crossings, which emerge during variations of $\lambda$, could be utilized to identify nontrivial higher-order phases. For a D-class system, it suffices to consider zero-energy crossings on the high-symmetry lines, namely $K=0$ and $\pi$, in the $(k,\lambda)$ parameter space. The higher-order topology is characterized by a pair of $\mathbb{Z}_2$ invariants, denoted as $(\mathcal{M}_x, \mathcal{M}_y)$, which relates to the total count of zero-energy crossings. When either of these invariants assumes the value $-1$, the system is anticipated to harbor Majorana corner states, as depicted in Fig. \ref{fig2}(a). In this article, we focus on the DIII and BDI symmetry classes, and aim to demonstrate that, zero-energy crossings emerging at any point in the parameter space may significantly influence the higher-order topology in these classes. The topological invariants are determined by the cumulative topological charges of all crossings.

\section{DIII class}\label{sec:III}

In the tenfold classification scheme \cite{chiu2016}, a DIII-class system is characterized by three intrinsic symmetries, namely time-reversal ($T$), particle-hole ($P$) and chiral symmetry ($S$). They act on the Hamiltonian $H_a(k,\lambda)$ according to
\begin{align}
    &T H_a(k,\lambda) T^{-1} = H_a(-k,\lambda),\ \ T^2=-1 \nonumber \\
    &P H_a(k,\lambda) P^{-1} = -H_a(-k,\lambda),\ \ P^2=1 \nonumber \\
    &S H_a(k,\lambda) S^{-1} = -H_a(k,\lambda),\ \ S^2=1 \label{eq:symmetry}
\end{align}
where $T$ and $P$ are anti-unitary operators and $S$ is a unitary operator. From Eq.(\ref{eq:symmetry}) we observe that time-reversal and particle-hole transformations act as reflections in the $(k,\lambda)$ parameter space, with the high-symmetry lines at $K=0,\pi$ serving as the mirror lines. As a result, each crossing at $(k_F,\lambda_F)$ has its partner at $(-k_F,\lambda_F)$. Additionally, chiral symmetry enforces that zero energy levels at any point in the parameter space are degenerate, with the degeneracy being twice the number of zero-energy crossings at that point. On the high-symmetry lines, if zero-energy crossings exist, they are anticipated to come in pairs due to Kramers degeneracy, thereby ensuring that the total count of zero-energy crossings on these lines is always even. In D-class systems, the parity of this number is closely related to the higher-order topological invariant $\mathcal M_a$. The emergence of Majorana corner states is expected when the parity is odd, i.e., $\mathcal M_x$ or $\mathcal M_y$ equals $-1$, as illustrated in Fig. \ref{fig2}(a). Apparently, these invariants always take trivial values in DIII-class systems, and hence could not be used to characterize the higher-order topology therein. In our subsequent discussions, we introduce a pair of $\mathbb Z_2$ invariants, which also relate to zero-energy crossings, to diagnose the nontrivial higher-order topology of the DIII class.

\subsection{$\mathbb Z_2$ invariants}

Following the same procedure as in the D class, we relate the higher-order topology in the 2D system to the first-order topology of the corresponding cylindrical Hamiltonian $\mathcal H_a(0)$, which describes a quasi-1D system with PBC in the $\bar a$ direction. Due to Kramers degeneracy, Majorana zero modes are always paired in real space, forming Majorana Kramers pairs. When such pairs manifest at two of the four corners in a square sample, as shown in Figs. \ref{fig1} and \ref{fig2}(b), we can expect the quasi-1D system, extending along the $x$ or $y$ direction, to exhibit nontrivial first-order topology. In DIII class, the first-order topology is characterized by topological invariants of $\mathbb Z_2$ type.

To further our analysis, we note that the Bloch BdG Hamiltonian $H_a(k,\lambda)$ can be transformed into an off-diagonal form in the eigenbasis of the chiral symmetry operator. This is represented by
\begin{equation}
    U_S^\dagger H_a(k,\lambda)U_S=
    \begin{pmatrix}
        0 & D_a(k,\lambda)  \\
        D_a^\dagger (k,\lambda) & 0
    \end{pmatrix},\label{eq:off_diag}
\end{equation}
where $U_S$ is the unitary matrix that diagonalizes chiral symmetry operator $S$. For the DIII class, with an appropriate selection of $U_S$, the $D$ matrix in Eq.(\ref{eq:off_diag}) can always be written in a form that satisfies $D_a(k,\lambda)=-D_a^T(-k,\lambda)$ \cite{schnyder2011}, implying that the $D$ matrix on high-symmetry lines, $D_a(K,\lambda)$, is antisymmetric. Employing singular value decomposition, we write $D_a(k,\lambda) = U_a(k,\lambda)\Lambda_a(k,\lambda) V_a^\dagger(k,\lambda)$, where $U_a$ and $V_a$ are both unitary matrices, and $\Lambda_a$ is a diagonal matrix with nonnegative entries. We then define a unitary $q$ matrix, $q_a(k,\lambda)=U_a(k,\lambda)V_a^\dagger(k,\lambda)$. The first-order topology of the quasi-1D Hamiltonian $\mathcal H_a(\lambda)$ is characterized by topological invariant $\tilde \nu_a(\lambda)$ in terms of the $q$ matrix \cite{qi2010}, which reads
\begin{align}
    (-1)^{\tilde\nu_a(\lambda)} = & \frac{\text{Pf}[q_a(0,\lambda)]}{\text{Pf}[q_a(\pi,\lambda)]}\frac{\sqrt{\det[q_a(\pi,\lambda)]}}{\sqrt{\det[q_a(0,\lambda)]}} = \frac{\text{Pf}[q_a(0,\lambda)]}{\text{Pf}[q_a(\pi,\lambda)]} \nonumber\\
    \times & \exp\left\{\frac{1}{2}\int_{0}^{\pi}dk\partial_{k} \ln\det[q_a(k,\lambda)]\right\}, \label{eq:top_inv_DIII_q} 
\end{align}
where ``Pf" represents Pfaffian. In Eq.(\ref{eq:top_inv_DIII_q}), $\sqrt{\det[q_a(k,\lambda)]}$ is required to be in the same branch for $k\in[0,\pi]$. Notably, $\tilde\nu_a(0)$ and $\tilde\nu_a(1)$ represent the first-order topological invariants of cylindrical and toroidal systems, respectively. We denote $\nu_a = \tilde \nu_a(0)$, and the pair of invariant, $(\nu_x,\nu_y)$, can precisely characterize nontrivial higher-order phases. Specifically, when either $\nu_x$ or $\nu_y$ equals 1, the 2D system is expected to host two Majorana Kramers pairs distributed at two corners, as shown in Fig. \ref{fig2}(b). Alternatively, utilizing the two equalities, $\text{Pf}(D_a) = \sqrt{|\det (D_a)|}\text{ Pf}(q_a)$ and $\det(D_a) = |\det (D_a)| \det(q_a)$, the topological invariants can also be expressed using the $D$ matrix, which reads
\begin{align}
    (-1)^{\tilde\nu_a(\lambda)} = & \frac{\text{Pf}[D_a(0,\lambda)]}{\text{Pf}[D_a(\pi,\lambda)]}\frac{\sqrt{\det[D_a(\pi,\lambda)]}}{\sqrt{\det[D_a(0,\lambda)]}} = \frac{\text{Pf}[D_a(0,\lambda)]}{\text{Pf}[D_a(\pi,\lambda)]} \nonumber\\
    \times & \exp\left\{\frac{1}{2}\int_{0}^{\pi}dk\partial_k \ln\det[D_a(k,\lambda)]\right\}.\label{eq:top_inv_DIII_D}
\end{align}

Equations (\ref{eq:top_inv_DIII_q}) and (\ref{eq:top_inv_DIII_D}), both formulated using cylindrical Hamiltonian $H_a(0)$, do not elucidate the link between corner modes and bulk Hamiltonian. To achieve this, our subsequent analysis aims to connect the pair of invariants to the zero-energy crossings that occur as the parameter $\lambda$ varies. These crossings can be derived from bulk Hamiltonian of the 2D system, thus allowing us to establish a direct bulk-boundary correspondence in higher-order phases.

\begin{figure}
    \includegraphics[width=0.48\textwidth]{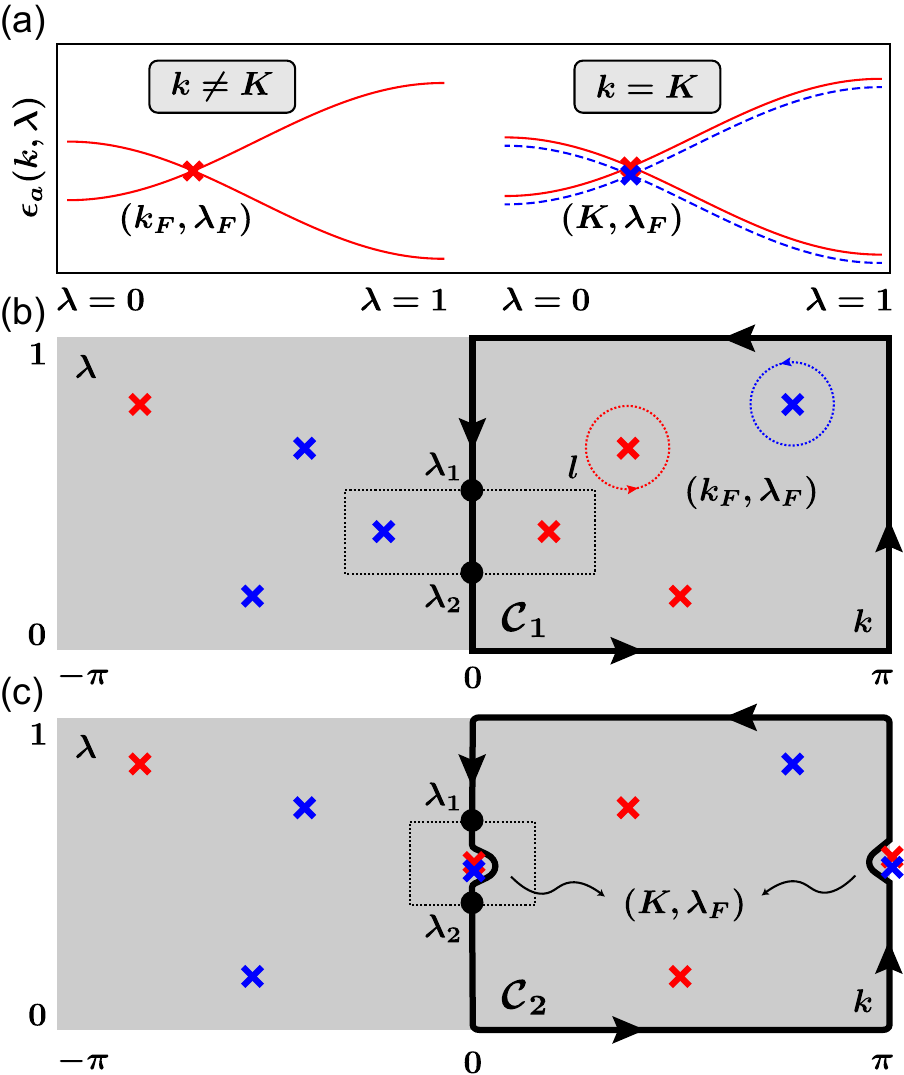}
    \caption{Zero-energy crossings on and off high-symmetry lines in the DIII class. (a) Variations of energy spectra with $\lambda$ for fixed $k$. On high-symmetry lines (right panel), Kramers degeneracy requires all energy levels to be degenerate and hence zero-energy crossings always come in pairs, whereas there is no such restriction for the crossings off high-symmetry lines (left panel). The red (solid) and blue (dashed) lines are deliberately shifted for demonstration of the degeneracy. (b) The first case where the crossings only occur off high-symmetry lines. (c) The second case with crossings on high-symmetry lines. The loop $\mathcal C_2$ now excludes these crossings. The red and blue crosses indicate zero-energy crossings with opposite topological charges. A $\mathbb Z_2$ topological charge can be associated to each pair of crossings, by considering a loop (black dashed lines in (b) and (c)) that encloses the pair of crossings.}\label{fig3}
\end{figure}

\subsection{Zero-energy crossings}

Before delving into the relationship between zero-energy crossings and higher-order topological invariants, we first take a look at the physical meaning of $\tilde\nu_a(1)$, which is associated with the torus Hamiltonian $\mathcal H_a(1)$. When interpreting the torus system as a quasi-1D system along the $\bar a$ direction, a nontrivial $\tilde\nu_a(1)$ implies the presence of Majorana Kramers pairs at both ends. This translates to the existence of gapless Majorana modes propagating along the edges in the $a$ direction of the 2D system, a feature of first-order topological superconductors in the DIII class. However, our interest lies in topological superconductors that exhibit nontrivial higher-order topology while maintaining trivial first-order topology, characterized by gapped edge spectra. Therefore, in our context, $\tilde\nu_a(1)$ is always assumed to be zero. This suggests that, in the nontrivial higher-order phase, the $\mathbb Z_2$ invariant $\tilde\nu_a(\lambda)$ is expected to change at some point as $\lambda$ is continuously varied from 1 to 0. This transition point is exactly where zero-energy crossings occur. In this regard, zero-energy crossings reveal the topological distinctions between the toroidal and the cylindrical system, which is crucial to our understanding of the bulk-boundary correspondence in higher-order topological phases.

Combining the expression of $\tilde\nu_a(\lambda)$ for $\lambda=0$ and $\lambda=1$ from Eq.(\ref{eq:top_inv_DIII_D}), we arrive at an equivalent formula for the higher-order topological invariant $\nu_a$, namely $\tilde\nu_a(0)$, which is given by
\begin{align}
    (-1)^{\nu_a} = & \frac{\text{Pf}[D_a(0,0)]}{\text{Pf}[D_a(0,1)]}\frac{\text{Pf}[D_a(\pi,1)]}{\text{Pf}[D_a(\pi,0)]} \nonumber \\ 
    \times & \exp\left\{\frac{1}{2}\int_{0}^{\pi}dk\partial_{k} \ln\frac{\det[D_a(k,0)]}{\det[D_a(k,1)]}\right\}. \label{eq:top_inv_DIII_D_1}
\end{align}
Our next step is to show that the ratio of Pfaffian terms in Eq.(\ref{eq:top_inv_DIII_D_1}) can be reformulated as an integral. It is important to note that while the $D$ matrix on high-symmetry lines ($K=0$ or $\pi$), denoted as $D_a(K,\lambda)$, is well defined for all $\lambda\in [0,1]$, its determinant may become zero. Following this observation, our subsequent analysis will distinguish between two cases.

In the first case, we assume $\det[D_a(K,\lambda)]\neq 0$ for all values of $\lambda$, indicating that there are no zero-energy crossings on the high-symmetry lines. We then select $N_\lambda$ equally spaced points along these lines, with $\lambda_n = n/N_\lambda$ for integers $n$ in the range $[0,N_\lambda]$. This allows us to express the ratio of Pfaffians in Eq.(\ref{eq:top_inv_DIII_D_1}) as the product of ratios for adjacent points, which reads
\begin{align}
    \frac{\text{Pf}[D_a(K,1)]}{\text{Pf}[D_a(K,0)]} &= \prod_{n=0}^{N_\lambda-1} \frac{\text{Pf}[D_a(K,\lambda_{n+1})]}{\text{Pf}[D_a(K,\lambda_n)]}. \label{eq:Pfaffian1}
\end{align}
Under the condition that $\det[D_a(K,\lambda)]\neq 0$, the phase of $\text{Pf} [D_a(K,\lambda)]$ can be made to vary continuously with $\lambda$, and so is the phase of $\sqrt{\det[D_a(K,\lambda)]}$. Due to this continuity we have
\begin{align}
    \frac{\text{Pf}[D_a(K,\lambda_{n+1})]}{\text{Pf}[D_a(K,\lambda_n)]} = \frac{\sqrt{\det[D_a(K,\lambda_{n+1})]}}{\sqrt{\det[D_a(K,\lambda_n)]}}.\label{eq:PfDet}
\end{align}
With this equality, we may further express Eq.(\ref{eq:Pfaffian1}) in an integral form in the limit of infinitely large $N_\lambda$, given by
\begin{align}
    \frac{\text{Pf}[D_a(K,1)]}{\text{Pf}[D_a(K,0)]} = \exp \left\{ \frac{1}{2} \int_0^1 d\lambda \partial_\lambda \ln\det[D_a(K,\lambda)] \right\}.\label{eq:Pfaffian}
\end{align}
From Eq.(\ref{eq:Pfaffian}) it follows that the ratio of Pfaffian in the left side of Eq.(\ref{eq:Pfaffian}) is related to line integrals of the determinant of $D$ matrix along high-symmetry lines, similar to the integral in Eq.(\ref{eq:top_inv_DIII_D}), which is calculated along lines with a constant $\lambda$ in the parameter space. Substituting Eq.(\ref{eq:Pfaffian}) into Eq.(\ref{eq:top_inv_DIII_D_1}), we obtain
\begin{align}
    (-1)^{\nu_a} = &\exp \left\{ \frac{1}{2} \oint_{\mathcal C_1} dl \nabla_l \ln \det[D_a(k,\lambda)] \right\} \nonumber \\
    = & \exp\left[i\pi\sum_{(k_F,\lambda_F)}n_{a}(k_F,\lambda_F) \right], \label{eq:top_inv_DIII_1}
\end{align}
where the integration along the loop $\mathcal C_1$, illustrated in Fig. \ref{fig3}(a), is proportional to the winding number of the $D$ matrix. In the second equality of Eq.(\ref{eq:top_inv_DIII_1}), we identify the winding number as the sum of topological charges for all zero-energy crossings $(k_F,\lambda_F)$ enclosed by $\mathcal C_1$, with $n_a(k_F,\lambda_F)$ representing the corresponding charge. In practice, $n_a$ is determined by calculating the winding number along a loop that encloses the crossing, such as the circle $l$ in Fig. \ref{fig3}(b). Time-reversal symmetry guarantees that each crossing at $(k_F,\lambda_F)$ has its partner at $(-k_F,\lambda_F)$. Due to the relation $D_a(k_F,\lambda_F) = -D_a(-k_F,\lambda_F)^T$, the two partners have opposite charges, i.e., $n_a(k_F,\lambda_F) = -n_a(-k_F,\lambda_F)$. Thus, the total topological charges of zero-energy crossings on either half of the parameter space could give the same $\nu_a$.

Now let us consider the second case, where zero-energy crossings also occur on high-symmetry lines. Considering that Pfaffian of the $D$ matrix becomes zero at the crossing $(K,\lambda_F)$, we need to exclude them while selecting the sample points in Eq.(\ref{eq:Pfaffian1}). Also, Eq.(\ref{eq:PfDet}) may fail if there exists crossings between the two adjacent points, which we denote as $(K,\lambda_F\pm\delta \lambda)$. The phase difference of the Pfaffian between these two points is not necessarily infinitely small while $\delta\lambda$ approaches zero. In this case, the loop $\mathcal C_1$ in Eq.(\ref{eq:top_inv_DIII_1}) is replaced by $\mathcal C_2$ shown in Fig. \ref{fig3}(b), which exactly circumvents the crossings. All the sample points still lie on the high-symmetry lines. On the vertical line segments of $\mathcal C_2$ , $\sqrt{\det[D_a(k,\lambda)]}$ can be made to vary continuously, thus ensuring $\sqrt{\det[D_a(K,\lambda_F+\delta \lambda)]}/\sqrt{\det[D_a(K,\lambda_F-\delta \lambda)]}$ to be infinitely small. This leads to the relation
\begin{align}
    \frac{\text{Pf}[D_a(K,\lambda_F+\delta \lambda)]}{\text{Pf}[D_a(K,\lambda_F-\delta \lambda)]} & = (-1)^{\eta_{a}(K,\lambda_F)} \label{eq:eta}\\
    & \times \frac{\sqrt{\det[D_a(K,\lambda_F+\delta\lambda)]}}{\sqrt{\det[D_a(K,\lambda_F-\delta\lambda)]}}, \nonumber
\end{align}
where $\eta_a(K,\lambda_F)$ is an integer that depends on the crossing at $(K,\lambda_F)$. Comparing with Eq.(\ref{eq:PfDet}), there is an additional phase factor in Eq.(\ref{eq:eta}). An odd $\eta_a$ suggests a $\pi$-phase shift in $\text{Pf}[D_a(K,\lambda)]$ across the zero-energy crossings.

In order to determine the precise value of $\eta_a$, we decompose the $2N\times 2N$ antisymmetric matrix $D_a(K,\lambda)$ into the form $D_a = Q_a^T M_a Q_a$, with $Q_a$ being a unitary matrix, and $M_a$ taking the block diagonal form as follows,
\begin{equation}
    M_a = \text{diag} \left\{ \begin{pmatrix} 0& m_{a,1} \\ -m_{a,1} & 0 \end{pmatrix}, ..., \begin{pmatrix} 0& m_{a,N} \\ -m_{a,N}& 0 \end{pmatrix} \right\}.\label{eq:M_a}
\end{equation}
For a non-singular $D$ matrix, $\pm m_{a,j}$, which represent eigenvalues of Hamiltonian $\mathcal H_a(K,\lambda)$, all take finite values. Assuming $|m_{a,1}|$ to be the smallest nonnegative energy level near the crossing $(K,\lambda_F)$, we then have $m_{a,1}=0$ exactly at the crossing. In the case of linear band crossing, we have $m_{a,1}\propto (\lambda-\lambda_F)$ near the crossing. As $\lambda$ crosses $\lambda_F$, a level crossing occurs between $m_{a,1}$ and $-m_{a,1}$, as Fig. \ref{fig3}(a) shows in the energy spectrum, and hence $m_{a,1}$ is expected to change its sign. We should note that, there are two degenerate crossings in the energy spectrum due to Kramers degeneracy, but $\pm m_{a,j}$ in Eq.(\ref{eq:M_a}) only account for half of the energy levels. While $\lambda$ crosses $\lambda_F$, $Q_a$ --- specifically, the phase of its determinant --- changes continuously. Alternatively, one may require $m_{a,1}$ to remain nonnegative, by swapping the positions of $\pm m_{a,1}$ in Eq.(\ref{eq:M_a}) after crossing $\lambda_F$, albeit at the expense of altering the sign of $\det[Q_a]$. Here, to clarify how zero-energy crossings affect the Pfaffian of $D$ matrix, we require $\det[Q_a]$ to be a continuous function of $\lambda$ near the crossing. By expressing the Pfaffian of $D$ matrix as
\begin{equation}
    \text{Pf}[D_a(K,\lambda)] = \det[Q_a(K,\lambda)] \prod_{j=1}^N m_{a,j}(K,\lambda),
\end{equation}
we readily find that the sign change of $m_{a,1}$ leads to an abrupt $\pi$-phase shift in the Pfafian of $D$ matrix. At $\lambda_F$, it's possible for more than one pair of $\pm m_{j,a}$ to cross zero energy. The number of such pairs is precisely half that of zero singular values of $D_a(K,\lambda_F)$ if we only consider linear band crossings. According to Eq.(\ref{eq:eta}), each pair contributes to a $\pi$-phase shift, and hence this number is also equal to $\eta_a(K,\lambda_F)$. Since only the parity of $\eta_a$ matters, we may define the parity of $\eta_a$ to be the $\mathbb Z_2$ charge of the crossing pair on the high-symmetry lines. In this case, Eq.(\ref{eq:Pfaffian1}) still holds, as the sample points $\lambda_n$ do not include the crossing points. For the two sample points adjacent to the crossing, Eq.(\ref{eq:PfDet}) needs to be replaced by Eq.(\ref{eq:eta}). It then follows from Eq.(\ref{eq:top_inv_DIII_D_1}) that, the higher-order topological invariant in the second case is given by
\begin{align}
    (-1)^{\nu_a}  & = \prod_{(K,\lambda_F)}(-1)^{\eta_a(K,\lambda_F)} \nonumber \\
    & \times \exp\left[i\pi\sum_{(k_F\neq K,\lambda_F)}n_a(k_F,\lambda_F) \right].\label{eq:top_inv_DIII}
\end{align}
Comparing with Eq.(\ref{eq:top_inv_DIII_1}), Eq.(\ref{eq:top_inv_DIII}) also includes contributions from zero-energy crossings on high-symmetry lines. The summation in the second line of Eq.(\ref{eq:top_inv_DIII}) involves crossings in either half of the parameter space.

In both cases aforementioned, the pair of higher-order topological invariants $(\nu_x,\nu_y)$ are determined by the topological charges of zero-energy crossings in the parameter space. For the crossings off the high-symmetry lines, their topological charges, $n_a$, are of $\mathbb Z$ type, in contrast to the $\mathbb Z_2$ charge, namely $\eta_a$, for crossings on the high-symmetry lines. Although $n_a$ may take any integer values, the topological invariants only depend on its parity, as Eqs.(\ref{eq:top_inv_DIII_1}) and (\ref{eq:top_inv_DIII}) suggest. Besides, the time-reversal partner of each crossing actually doesn't provide additional information. With these observations, we may treat each pair of zero-energy crossings at $(\pm k_F,\lambda_F)$ as a whole, and associate a $\mathbb Z_2$ charge to them in a unified fashion, regardless of whether they appear on or off the high-symmetry lines. The topological charge, denoted as $\tilde n_a(k_F,\lambda_F)$, is defined according to \cite{schnyder2011}
\begin{equation}
    (-1)^{\tilde n_a(k_F,\lambda_F)} = \frac{\text{Pf}[D_a(K,\lambda_1)]}{\text{Pf}[D_a(K,\lambda_2)]}\frac{\sqrt{\det[D_a(K,\lambda_2)]}}{\sqrt{\det[D_a(K,\lambda_1)]}},\label{eq:z2charge}
\end{equation}
where the two points, $(K,\lambda_1)$ and $(K,\lambda_2)$, are the intersections of the high-symmetry line and the dashed square loop, as depicted in Figs. \ref{fig3}(b) and (c). The loop encloses only the pair of zero-energy crossings considered. In Eq.(\ref{eq:z2charge}), $\sqrt{\det [D_a (k,\lambda)]}$ is required to vary continuously along the left or right half of the loops, instead on the high-symmetry lines, where zero-energy crossings may appear. Equation (\ref{eq:z2charge}) reduces to Eq.(\ref{eq:eta}) when we consider an infinitely small loop and set $\lambda_{1}=\lambda+\delta \lambda$, $\lambda_{2}=\lambda-\delta \lambda$. Winding numbers of the $D$ matrix along these loops are always zero, but the $\mathbb Z_2$ charge defined in Eq.(\ref{eq:z2charge}) can be nontrivial. To be specific, $\tilde n_a(k_F,\lambda_F)=1$ if $\eta_a(K,\lambda_F)$ (for crossings on the high-symmetry lines) or $n_a(k_F,\lambda_F)$ (for crossings off high-symmetry lines) is odd; otherwise $\tilde n_a(k_F,\lambda_F)=0$. Consequently, the higher-order topological invariant $\nu_a$ simply reflects the parity of the total $\mathbb Z_2$ charges for all pairs of zero-energy crossings, i.e.
\begin{equation}
    \nu_a = \sum_{(k_F\geq 0,\lambda_F)}\tilde n_a(k_F,\lambda_F) \text{ mod }2.\label{eq:top_inv_DIII_z2}
\end{equation}

So we have established the connection between Majorana corner modes appearing in nontrivial higher-order phases and zero-energy crossings occurring while the toroidal system is deformed into the cylindrical one. To gain an intuitive understanding of this connection, we note that the higher-order topology we are studying in this work represents the topological difference (in the first-order sense) between the toroidal Hamiltonian $\mathcal H_a(\lambda_a=1)$ and the cylindrical Hamiltonian $\mathcal H_a(\lambda_a=0)$, both viewed as quasi-1D Hamiltonian. While $\lambda_a$ varies continuously in the range $[0,1]$, the first-order topological invariant of $\mathcal H_a(\lambda_a)$ only changes when the energy gap closes for certain value of $\lambda_a$, showing as zero-energy crossings in the $(k_{\bar a},\lambda_a)$ parameter space. The existence of a single crossing does not necessarily imply that two Hamiltonian $\mathcal H_a(\lambda_a)$ with distinct $\lambda_a$ on either side of the crossing are topologically distinct. We need additionally consider the topological charge of the crossing, which indicates its stability. A zero-energy crossing with nontrivial charge suggests that $\mathcal H_a(\lambda_a)$ cannot be smoothly deformed across it, thereby signaling a change in the topology of the quasi-1D Hamiltonian. The higher-order topology is then determined by all the (pairs of) crossings in the $(k_{\bar a},\lambda_a)$ parameter space, as indicated by Eq.(\ref{eq:top_inv_DIII}) or (\ref{eq:top_inv_DIII_z2}).

\begin{figure*}[t]
    \includegraphics[width=0.99\textwidth]{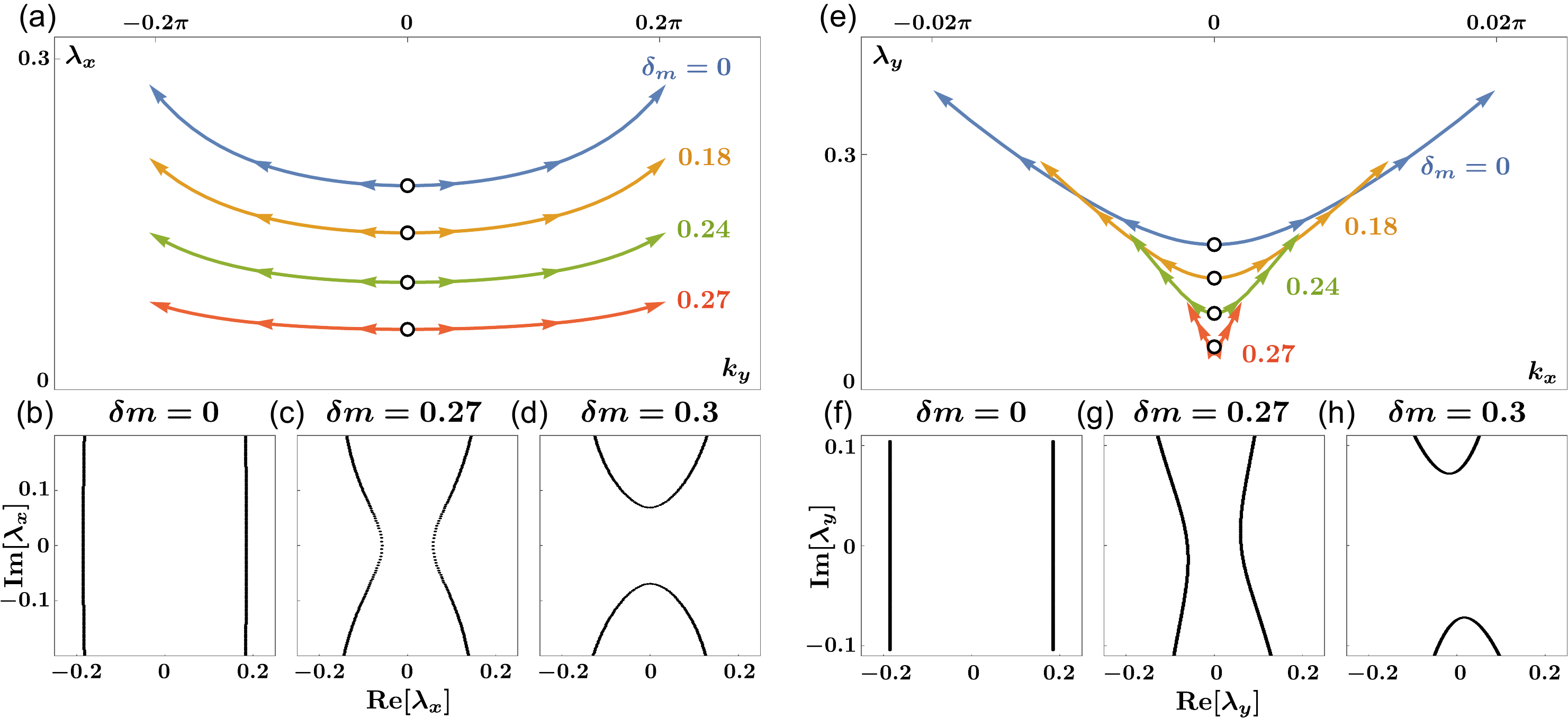}
    \caption{The flow of zero-energy crossings in the parameter space under the variations of $\delta m$ and $m_1$. (a) In the $(k_y,\lambda_x)$ space, a pair of crossings, marked as black circles, reside on the high-symmetry line $K=0$ when $m_1=0$. Finite $m_1$ separates the pair and drive them off the line in opposite directions, as indicated by the arrows. With the increase of $\delta m$, the crossings move towards $\lambda=0$ and vanish. (b)-(d) $\lambda$ profiles generated by solving Eq.(\ref{eq:detA}) for $k\in [-\pi,\pi)$. For sufficiently large $\delta m$, the characteristic equation has no real root, signaling the disappearance of zero-energy crossings. (e)-(h) A similar phenomenon is observed in the $(k_x,\lambda_y)$ space. The other parameters are set as $\mu=3t$, $\Delta=t$, $m=0.4t$, $\theta=\pi/4$, which, unless otherwise specified, remain the same in subsequent plots.}\label{fig4}
\end{figure*}

\subsection{Bulk-boundary correspondence}

Having related the zero-energy crossings to higher-order topology, we now demonstrate how these crossings and their topological charges are effectively identified from the bulk Hamiltonian, so as to make a direct bulk-boundary correspondence in higher-order phases. This approach is applicable to systems in both DIII and BDI classes.

To begin with, we note that at the crossing point $(k_F,\lambda_F)$, $\lambda_F$ is a root of the characteristic equation $\det[H_a(k,\lambda)]=0$ \cite{rhim2018}. Its multiplicity is even due to the presence of chiral symmetry. In terms of $D$ matrix, the characteristic equation reduces to
\begin{equation}
    \det[D_a(k,\lambda)]=0,\label{eq:chara_eq}
\end{equation}
except that the root's multiplicity is halved. In particular, for linear crossings on the high-symmetry lines, we have $\det[D_a(K,\lambda)]\propto (\lambda-\lambda_F)^{p_a}$ near the crossing, from which it follows that the root's multiplicity $p_a$ is equal to the number of zero singular values of $D_a(K,\lambda_F)$ ($\pm m_{a,1}$ in Eq.(\ref{eq:M_a})), which exactly equals $2\eta_a(K,\lambda_F)$, i.e., $p_a/2=\eta_a$. Hence, the root's multiplicity determines the $\mathbb Z_2$ charge of zero-energy crossings on the high-symmetry lines. It's possible that the crossings are of quadratic or cubic type, and there may exist a number of crossings with different types at the same point $\lambda_F$. We will discuss this generic case in Sec. \ref{sec:V}, where it is demonstrated that the $\mathbb Z_2$ charge is still given by the parity of $p_a/2$.

Following Eq.(\ref{eq:top_inv_DIII_1}), the topological charge for each individual crossing off the high-symmetry lines is equal to the winding number of the $D$ matrix along a loop that encloses the crossing. No matter where the crossings occur, we need to compute the determinant of the $D$ matrix. To demonstrate how it is obtained from the bulk Hamiltonian, we first write down $H_a(k,\lambda)$ as
\begin{equation}
    H_a(k,\lambda) = H_a(k,1) - (1-\lambda)B_a(k)\label{eq:ham_relation}
\end{equation}
in the Bloch basis $\Psi_a(k)$, with $B_a(k)$ being the Bloch Hamiltonian of boundary Hamiltonian $\mathcal B_a$ in Eq.(\ref{eq:ham}). We denote the components of $\Psi_a(k)$ as $[c_a(k)]_{i\alpha}$, where the Latin letter $i$ labels the unit cell along the $a$ direction, and the Greek letter $\alpha$ labels the degree of freedom within each cell. The entries of the Bloch Hamiltonian matrix are thus represented by $[H_a(k,\lambda)]_{i\alpha,j\beta}$. Since chiral symmetry operates within the inner space, the corresponding $D$ matrix has a similar relation as Eq.(\ref{eq:ham_relation}), which reads 
\begin{equation}
    D_a(k,\lambda) = D_a(k,1) A_a(k,\lambda),
\end{equation}
with
\begin{equation}
    A_a(k_,\lambda)=\mathbb I-(1-\lambda) D_a^{-1}(k,1) D^b_a(k).\label{eq:Aa}
\end{equation}
Here, $\mathbb I$ represents the identity matrix, and $D^b_a(k)$ is the $D$ matrix of $B_a(k)$.

For higher-order phases with gapped bulk spectrum, $\det[D_a(k,1)]$ always takes a finite value, and hence Eq.(\ref{eq:chara_eq}) is equivalent to 
\begin{equation}
    \det [A_a(k,\lambda)] = 0.\label{eq:chara_eqA}
\end{equation}
The boundary Hamiltonian $B_a(k)$ only involves boundary degrees of freedom, which we denote by $i_b\alpha$. Consequently, the only possible nonzero entries of $B_a(k)$ are $[B_a(k_{\bar a})]_{i_b\alpha,j_b\beta}$. As such, we demonstrate in Appendix that $\det[A_a(k,\lambda)]$ is equal to the determinant of a $r_a\times r_a$ matrix, denoted by $\tilde{A}_a(k,\lambda)$, with $r_a$ representing the rank of $D_a^b(k)$. The specific form of $\tilde{A}_a(k,\lambda)$ is given by
\begin{align}
    \tilde{A}_a(k_{\bar a},\lambda_a) & = \mathbb I_{r_a\times r_a} - \frac{1}{2\pi}(1-\lambda_a) \nonumber \\
     & \times \int dk_a \tilde V_a^{b\dagger} [F(k_{a}) \otimes D^{-1}(\bm k)] \bar D^b_a(k_{\bar a})\tilde V_a^b, \label{eq:tildeA}
\end{align} 
where we recover the subscripts of $k$ and $\lambda$ to avoid confusion. In Eq.(\ref{eq:tildeA}), $F(k_a)$ is related to Fourier transformation with $[F(k_a)]_{i_b,j_b} = e^{ik_a(i_b-j_b)}$, $D(\bm k)$ is the $D$ matrix of bulk Hamiltonian in 2D momentum space, and $\bar D^b_a(k_{\bar a})$ is the block that corresponds to boundary cells in matrix $D^b_a(k_{\bar a})$. $\tilde V_a^b$ is a semi-unitary matrix related to compact singular value decomposition of $\bar D^b_a(k_{\bar a})$.

As a result, Eq.(\ref{eq:chara_eqA}) reduces to
\begin{align}
    \det [\tilde{A}_a(k,\lambda)] = 0.\label{eq:detA}
\end{align}
The loop integral in Eq.(\ref{eq:top_inv_DIII}) may also be reexpressed in terms of $\tilde{A}_a(k,\lambda)$, by noting that with trivial first-order topology we always have
\begin{align}
    \oint_{\mathcal C} dl \nabla_l \ln \det[D_a(k,1)] = 0
\end{align}
for any loop $\mathcal C$ in the parameter space. The topological charge of each zero-energy crossing is then determined according to
\begin{align}
    n_a(k_F,\lambda_F) =  \frac{1}{2\pi i} \oint_{\mathcal C} dl \nabla_l \ln \det[\tilde A_a(k,\lambda)], \label{eq:top_inv_DIII_A}
\end{align}
with $\mathcal C$ being any loop that encloses the crossing $(k_F,\lambda_F)$ only. Therefore, all calculations involving the determinant of $D_a(k,\lambda)$ can now be replaced by that of $\tilde{A}_a(k,\lambda)$, which, according to Eq.(\ref{eq:tildeA}), is obtained from the $D$ matrix of bulk Hamiltonian, $D(\bm k)$. After finding all the zero-energy crossings and their topological charges, we immediately obtain the topological invariants according to Eq.(\ref{eq:top_inv_DIII}). 

Thus far, we have demonstrated that zero-energy crossings serve as a bridge connecting bulk Hamiltonian and Majorana corner modes. Although zero-energy crossings are determined from the bulk Hamiltonian, their numbers may change when bulk gap remains open. This happens when topological phases transitions are driven by the closure of edge gaps, as we shall demonstrate in the following example.

\subsection{An example}

We consider a spinful toy model in 2D with two orbital degrees in each unit cell. The Bloch BdG Hamiltonian takes the form
\begin{align}
    & H_{\text{DIII}}(\bm k)  = \epsilon_{\bm k}\tau_z + 2\Delta(\sin k_x\tau_y - \sin k_y \tau_x s_z \sigma_z) \label{eq:ham_DIII}\\
    & + m(\sin\theta \tau_z\sigma_x - \cos\theta s_z\sigma_y)  - \delta m\tau_x s_z \sigma_y - m_1\tau_x s_x \sigma_y,\nonumber 
\end{align}
where the kinetic energy $\epsilon_{\bm k} = \mu-2t(\cos k_x + \cos k_y)$. Here $\Delta$ represents the $p$-wave pairing amplitude, and the Pauli matrices $\tau$, $s$ and $\sigma$ act in particle-hole, spin and orbital space, respectively. The three symmetry operators defined in Eq.(\ref{eq:symmetry}) are represented as $T = -is_y \mathcal K$, $P = \tau_x \mathcal K$ and $S = \tau_x s_y$, where $\mathcal K$ denotes the complex conjugation operator. Additionally, the model preserves inversion symmetry when $m_1=\delta m=0$, with the corresponding operator being $\mathcal I = \tau_z$, up to a gauge factor.

The first two terms in Eq.(\ref{eq:ham_DIII}) describe two copies ($\sigma_z=\pm 1$) of topological superconductors with $p\pm ip$ pairing. Each copy features two counter-propagating helical Majorana modes on edges in the nontrivial phase. The three onsite terms $m$, $\delta m$ and $m_1$ can gap out the helical modes. In particular, the $m$ term introduces anisotropic mass gaps to the four edges, controlled by $\theta$. This is crucial for the emergence of Majorana corner modes, which appear when two adjacent edges exhibit mass gaps of opposite signs. In the absence of $m_1$, $s_z$ is a good quantum number, and the model simply reduces to two independent copies ($s_z = \pm 1$) of the D-class model we introduced in Ref.\cite{wang2023c}. In the D-class model, the nontrivial higher-order topology is captured by zero-energy crossings on high-symmetry lines. Crossings off these lines are not protected due to the absence of chiral symmetry. In contrast, for the DIII class, zero-energy crossings both on and off these lines contribute to the higher-order topology. Time-reversal symmetry-preserving perturbations, such as the $m_1$ term, either couple and eliminate the pair of crossings on high-symmetry lines or separate them away from these lines, depending on the $\mathbb Z_2$ charge of the pair defined in Eq.(\ref{eq:z2charge}).

\begin{figure}[t]
    \includegraphics[width=0.48\textwidth]{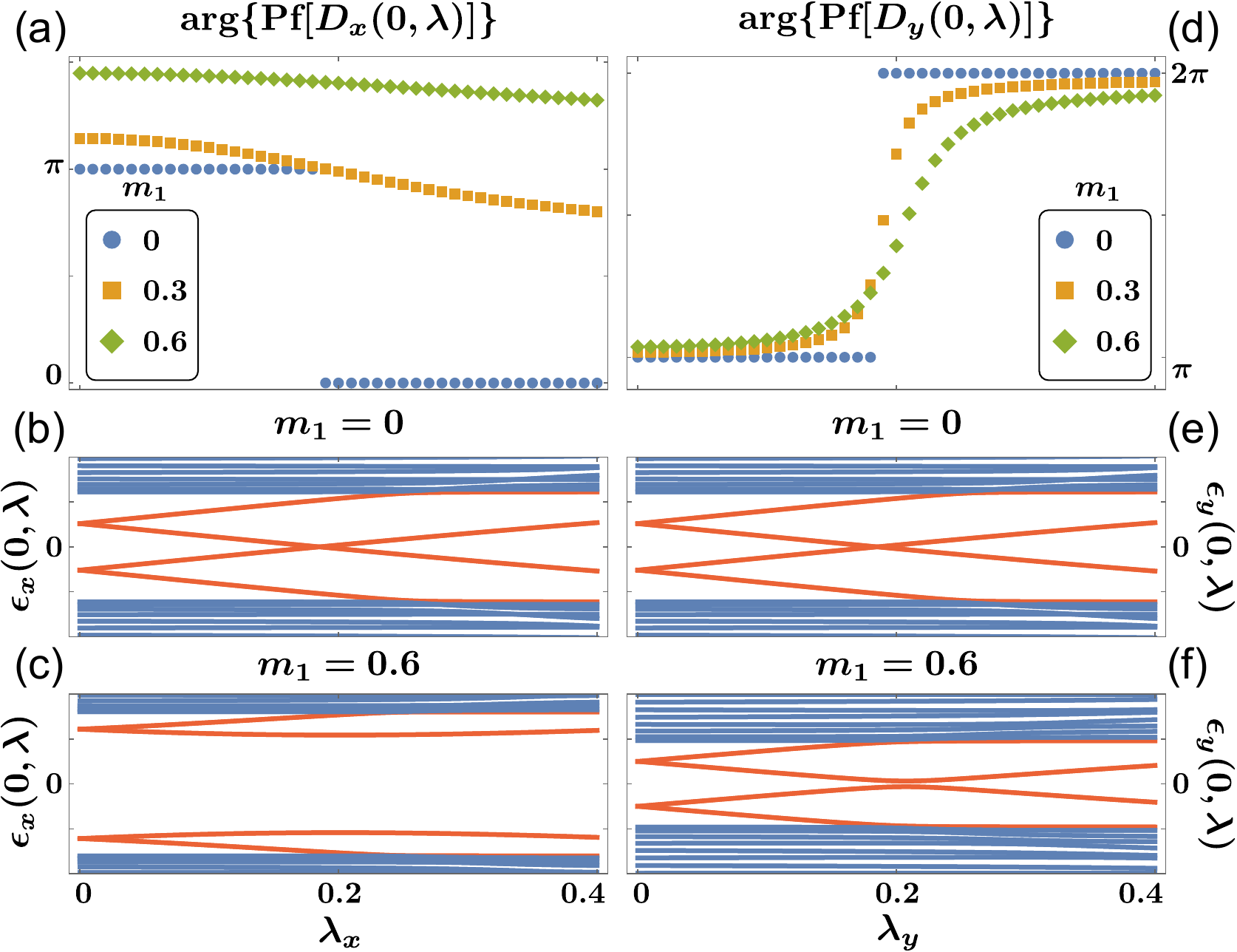}
    \caption{The $\pi$-phase jump in Pfaffian of the $D$ matrix induced by zero-energy crossings. (a), (d) zero-energy crossings occurring on high-symmetry lines when $m_1=0$ leads to an abrupt $\pi$-phase jump in $\text{Pf}[D_a(0,\lambda)]$. For cases with finite $m_1$, there is no such discontinuity. (b)-(c), (e)-(f) Energy spectra along $K=0$. No zero-energy crossing exists on the high-symmetry line when $m_1$ is finite. In these plots, $N_a$, the number of unit cells along the $a$ direction, is chosen to be 40. The Pfaffian is calculated using the code from Ref.\cite{wimmer2012}.}\label{fig5}
\end{figure}

Given that we focus on boundary terminations commensurate with unit cells, the boundary Hamiltonian is directly derived from the bulk Hamiltonian. In the boundary basis, which in this model includes only the first and last unit cells --- denoted as $\{\psi_{a,N_a}(k),\psi_{a,1}(k)\}$ with $N_a$ representing the number of unit cells in the $a$ direction and $\psi_{a,j}(k)$ encompassing all internal degrees of freedom $[c_a(k)]_{j\alpha}$ --- the resulting matrix $B_a$ takes the following simple form,
\begin{align}
    B_a = \begin{pmatrix} 0 & h_a \\ h_a^{\dagger} & 0 \end{pmatrix}
\end{align}
where $h_a$ includes all inter-cell hopping and pairing terms, with $h_x=-t\tau_z-i\Delta\tau_y$ and $h_y=-t\tau_z+ i\Delta \tau_x s_z \sigma_z$. The unitary matrix that sends the bulk Hamiltonian $H(\bm k)$ into block off-diagonal form is
\begin{align}
    U_S = \frac{1}{\sqrt{2}} \begin{pmatrix} s_0 & -is_y \\ s_y & is_0 \end{pmatrix}\otimes \sigma_0,
\end{align}
where $s_0$ and $\sigma_0$ are identity matrices in the spin and orbital space, respectively. The zero-energy crossings can then be identified according to Eqs.(\ref{eq:tildeA}) and (\ref{eq:detA}).

In Figs. \ref{fig4}(a) and (e), we plot the positions of zero-energy crossings in the parameter space for various values of $\delta m$ and $m_1$. When $m_1=0$, the system supports a pair of zero-energy crossings on the high-symmetry line $K=0$. The introduction of the $m_1$ term drives the pair of crossings away from this line along opposite directions. The arrows in Figs. \ref{fig4}(a) and (e) indicate where the zero-energy crossings flow with increasing $m_1$.

In this model, each zero-energy crossing at $(k_F,\lambda_F)$ is also paired with another one at $(k_F,-\lambda_F)$, the latter not explicitly shown here as we focus only on crossings with $\lambda_F\in [0,1]$. As $\delta m$ increases (see Figs. \ref{fig4}(a) and (e)), the crossings at $\pm\lambda_F$ migrate towards $\lambda=0$, eventually annihilating. A straightforward way to illustrate this transition is through the $\lambda$ profile, which displays all roots of Eq.(\ref{eq:detA}) for $k\in[-\pi,\pi)$. Zero-energy crossings correspond to real roots for $\lambda$. As depicted in Figs. \ref{fig4}(b)-(d) and (f)-(h), the zero-energy crossings vanish at sufficiently large $\delta m$, signaling edge phase transitions where Majorana zero modes at different corners may either change positions or couple. Our previous study in the D-class model \cite{wang2023c} showed that, in the absence of $m_1$, the energy gap on edges along the $x$ direction closes at $|\delta m| = |m \sin\theta|$, and along the $y$ direction at $|\delta m| = |m \cos\theta|$. The presence of $m_1$ term has a negligible impact on these gap-closing points. Pinpointing the phase transition position for finite $m_1$ is not our focus here. Instead, our next step is to examine the topological charges of these zero-energy crossings in order to derive the higher-order topological invariants.

\begin{figure}[t]
    \includegraphics[width=0.48\textwidth]{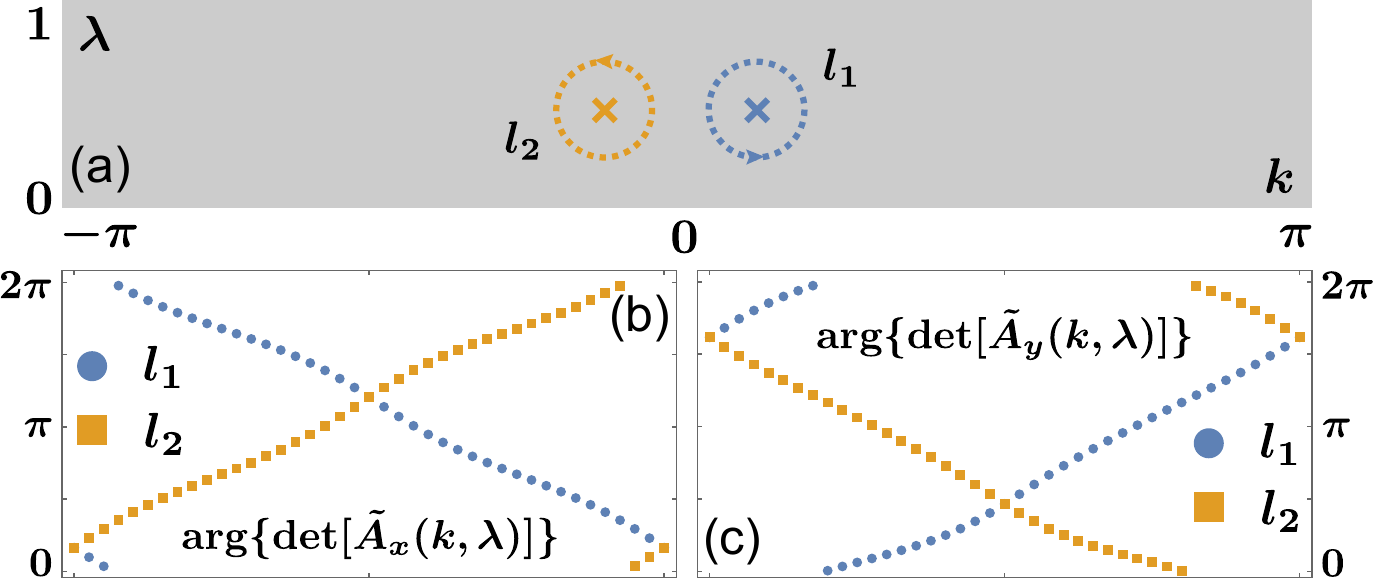}
    \caption{Topological charges of zero-energy crossings off high-symmetry lines. Nontrivial windings in the phase of $\det [\tilde{A}_a(k,\lambda)]$ are observed along the loops $l_1$ and $l_2$, each of which encircles a single zero-energy crossing. The two time-reversal partners at $\pm k_F$ carry opposite charges. In these plots, $m_1=0.3t$ and $\delta m=0$.}\label{fig6}
\end{figure}

\begin{figure}[t]
    \includegraphics[width=0.48\textwidth]{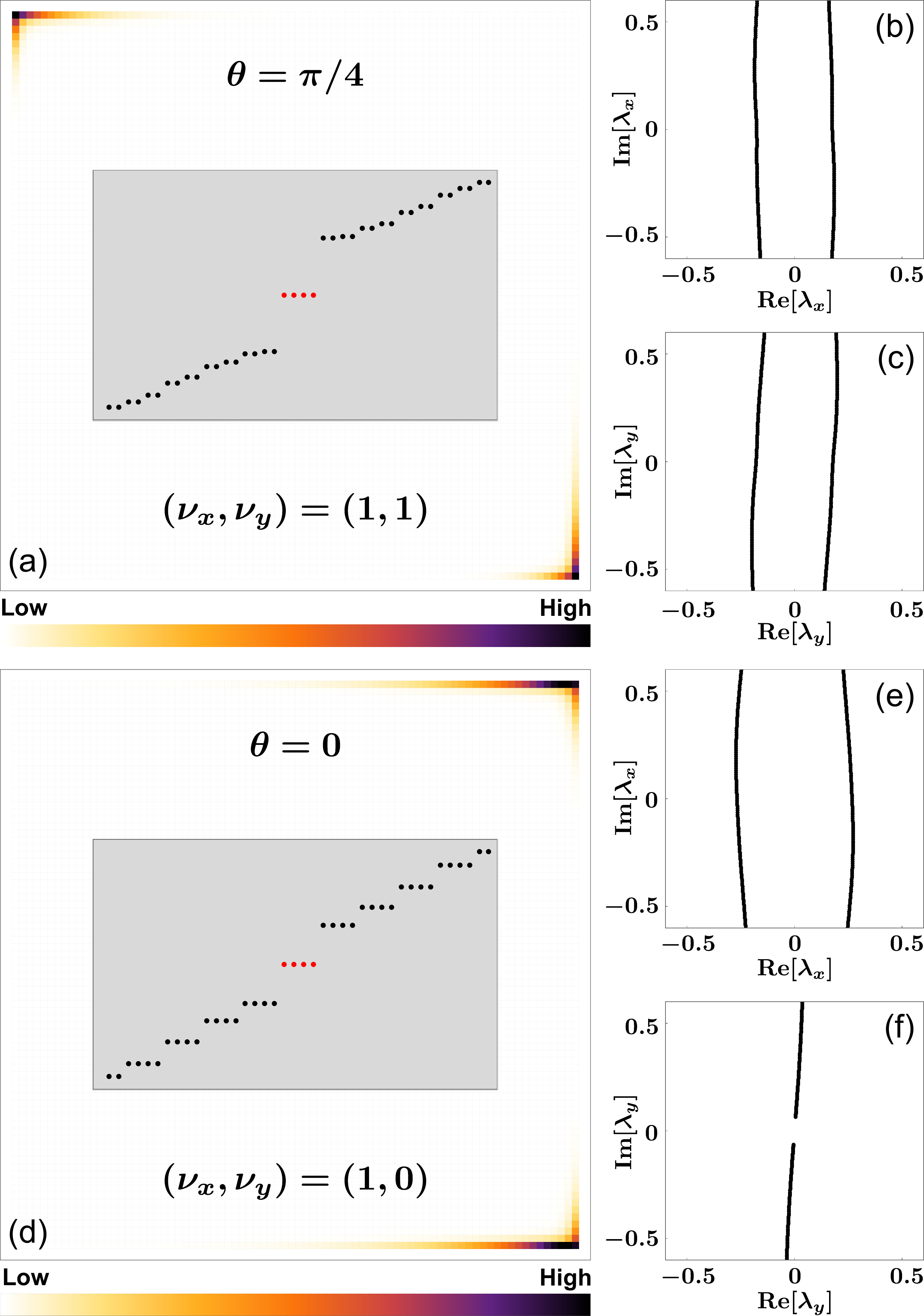}
    \caption{Distributions of Majorana Kramers pairs in nontrivial higher-order phases. (a) The probability distributions of Majorana zero modes. The inset shows the energy spectrum for an $80\times 80$ open-boundary system, with red dots representing Majorana zero modes. The two invariants, $\nu_x$ and $\nu_y$, both take nontrivial values, as indicated by $\lambda$ profiles in (b) and (c). This results in two Majorana Kramers pairs sitting at opposite corners. (d) Only one of the invariants is nonzero, as revealed by $\lambda$ profiles in (e) and (f), and the two Majorana pairs therefore reside on adjacent corners. In these plots, $m_1=0.3t$ and $\delta m=0.1t$.}\label{fig7}
\end{figure}

Let's first consider zero-energy crossings on high-symmetry lines, denoted as $(K,\lambda_F)$. We already know that $\lambda_F$ is a root of Eq.(\ref{eq:detA}) at $k=K$. According to Eq.(\ref{eq:eta}), $\eta_a(K,\lambda_F)$, which equals half the root's multiplicity, determines whether the Pfaffian of the $D$ matrix experiences a $\pi$-phase jump as $\lambda$ crosses this point. The plots corresponding to $m_1=0$ in Figs. \ref{fig5} (a) and (c) indeed exhibit such a jump, suggesting an odd $\eta_a$. By solving Eq.(\ref{eq:detA}), we obtain a real root with multiplicity 2, implying $\eta_a(K,\lambda_F)=1$. This is consistent with the energy spectrum along the high-symmetry line in Figs. \ref{fig5}(b) and (e), where each level is doubly degenerate due to Kramers degeneracy, resulting in exactly two linear band crossings at the zero energy for $m_1=0$. Turning on $m_1$ drives the crossings away from the high-symmetry line, and hence the Pfaffian of $D$ matrix varies continuously along the line, as seen in Figs. \ref{fig5}(a) and (d) for finite $m_1$. This leads us to the first case investigated in Sec. \ref{sec:III}B, where zero-energy crossings appear off the high-symmetry lines. 

Since the pair of crossings on the high-symmetry line has a nontrivial $\mathbb Z_2$ charge, we expect that finite $m_1$ won't couple them. To verify this argument, we may first locate the position of each crossing, followed by the determination of their topological charges according to Eq.(\ref{eq:top_inv_DIII_A}). Note that each crossing at $(k_F,\lambda_F)$ has its time-reversal partner at $(-k_F,\lambda_F)$, with opposite charge. In Figs. \ref{fig6}(b) and (c), we show the phase of $\det[\tilde A_a(k,\lambda)]$ along a contour that encircles each crossing for $m_1=0.3$, which clearly shows nontrivial windings, with topological charges being $\pm 1$. This suggests the $\mathbb Z_2$ charge associated with the pair of crossings remains $1$. Due to this nontrivial $\mathbb Z_2$ charge, finite $m_1$ can at most separate the two crossings away from each other.

Through the above analysis, we are able to obtain the pair of topological invariants $(\nu_x,\nu_y)$ as defined in Eq.(\ref{eq:top_inv_DIII}). For the examples shown in Figs. \ref{fig5} and \ref{fig6}, we have $(\nu_x,\nu_y)=(1,1)$, which indicates that two Majorana Kramers pairs sit at diagonals, as depicted in Fig. \ref{fig7}(a). When the inversion symmetry is enforced ($m_1=\delta m=0$), the two pairs always appear at opposite corners and never meet. This suggests that $\nu_x$ is always equal to $\nu_y$, and hence we can simply use one of them to characterize the higher-order phase protected by inversion symmetry. In this case, topological phase transition occurs only when the bulk gap closes. Without inversion symmetry, the two Majorana pairs may reside at any two corners, as shown in Fig. \ref{fig7}(d) for $\theta=0$, where zero-energy crossings only occur in $(k_y,\lambda_x)$ parameter space and hence $(\nu_x,\nu_y) = (1,0)$. It is possible that the two Majorana pairs in neighboring corners couple with each other through the closure of the edge gap, driving the system into trivial phases. Since the pair of invariants are protected by both bulk and edge gaps, they are able to indicate higher-order topology regardless of whether the inversion symmetry is enforced. When at least one of the two invariants is nonzero, we immediately know the system resides in the nontrivial higher-order phase, but not the other way around. For instance, if all four corners host Majorana Kramers pairs, both invariants would be zero. This argument also applies to the BDI class which we shall deal with in the following. 

\section{BDI class}\label{sec:IV}

\subsection{$\mathbb Z$ invariants}

\begin{figure}[t]
    \includegraphics[width=0.48\textwidth]{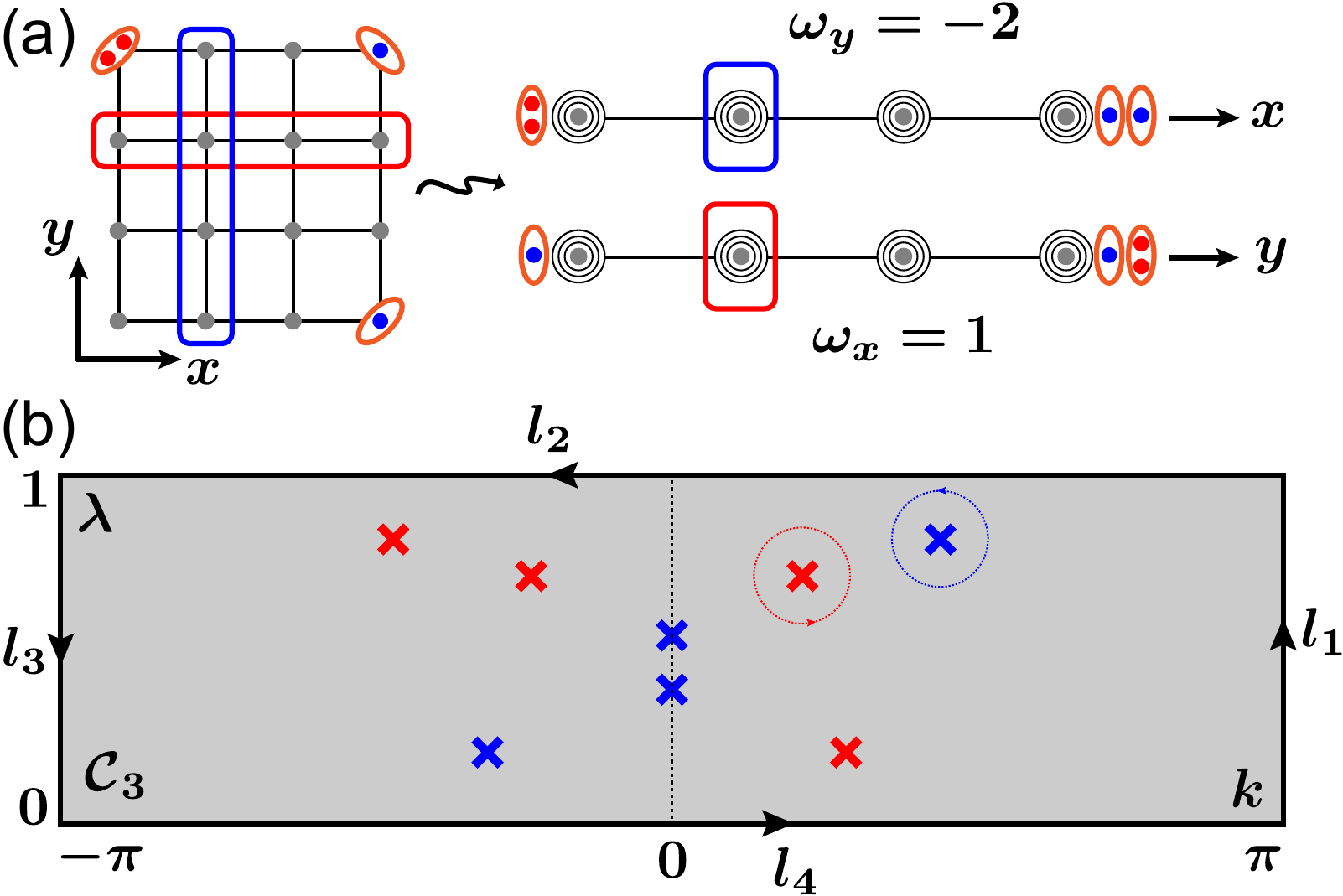}
    \caption{(a) The relation between higher-order topological invariants and Majorana corner modes in the BDI class. The invariant $\omega_y$ corresponds to the difference in the number of Majorana zero modes with opposite chiralities at two corners on the left, while $\omega_x$ counts the two bottom corners. Blue and red dots represent Majorana modes with chiralities $+1$ and $-1$, respectively. (b) Zero-energy crossings in the BDI class. The crossings off high-symmetry lines always come in pairs. On the line, the crossing may appear alone. In contrast to the DIII class, signs of topological charges for time-reversal partners, as indicated by colors of the crosses, can be identical.}\label{fig8}
\end{figure}

The Hamiltonian in the BDI class also satisfies the three intrinsic symmetries listed in Eq.(\ref{eq:symmetry}), except that the time-reversal operator has the property $T^2=1$. This key difference determines that zero-energy crossings on high-symmetry lines in the parameter space do not necessarily come in pairs, due to the absence of Kramers degeneracy, as schematically illustrated in Fig. \ref{fig8}(b). However, time-reversal symmetry still ensures that a crossing off the high-symmetry lines, say at $(k_F,\lambda_F)$, has a partner at $(-k_F,\lambda_F)$, only that the two may not have opposite topological charges, as the $D$ matrix does not satisfy the relation $D_a(k,\lambda)=-D^T_a(-k,\lambda)$. Hence each crossing within the pair must be examined individually. Consequently, the higher-order topological invariants in this case are of $\mathbb Z$ type.

In 2D, the BDI class does not have stable first-order topology but can exhibit nontrivial higher-order topology \cite{chiu2016}. Here, we continue to consider systems with gapped bulk and edges, meaning that Hamiltonian $H_a(k,\lambda)$ is gapped for both $\lambda=1$ and $\lambda=0$. We may associate a topological invariant to the boundary-modulated Hamiltonian for generic $\lambda$, akin to the $\tilde \nu_a(\lambda)$ in Eq.(\ref{eq:top_inv_DIII_D}) for the DIII class. The topological invariant in the BDI class is the winding number of $D$ matrix along the loop with constant $k$ and takes the form
\begin{equation}
    \tilde\omega_a(\lambda) = \frac{1}{2\pi i}\int_{-\pi}^{\pi}dk\partial_k \ln\det[D_a(k,\lambda)].\label{eq:winding_BDI}
\end{equation}
Similar to the DIII class, the higher-order topological invariant $\omega_a$ here is defined to be the first-order topological invariant of the quasi-1D cylindrical Hamiltonian $H_a(k,0)$, i.e., 
\begin{equation}
    \omega_a=\tilde\omega_a(0).\label{eq:topo_inv_BDI}
\end{equation}
Meanwhile, $\tilde\omega_a(1)$ indicates the first-order topology of a toroidal system and is hence always zero.

Following from Eqs.(\ref{eq:winding_BDI}) and (\ref{eq:topo_inv_BDI}), $\omega_a$ may take any integer value. In a 1D system, this invariant reflects the difference in the count of Majorana zero modes with distinct chiralities ($\pm 1$), namely the eigenvalues of chiral operator, at one end. For the Hamiltonian $H_a(k,\lambda)$ in consideration, we may denote $N_{a\pm}^{L(R)}$ as the number of Majorana zero modes at left (right) end with chirality $\pm 1$. Then we have 
\begin{equation}
    \omega_a=N_{a+}^L-N_{a-}^L=N_{a-}^R-N_{a+}^R.\label{eq:chirality}
\end{equation}
In a truly 1D system, Majorana zero modes of different chiralities can couple, resulting in only one type of Majorana modes residing at each end if no other symmetries are enforced. However, in the quasi-1D system here, each end, comprising two spatially separated corners, can accommodate stable Majorana corner modes of different types, as depicted in Fig. \ref{fig8}(a). The system enters nontrivial higher-order phases when one or both of the invariants $(\omega_x,\omega_y)$ take nonzero values.

\begin{figure*}[t]
    \includegraphics[width=0.99\textwidth]{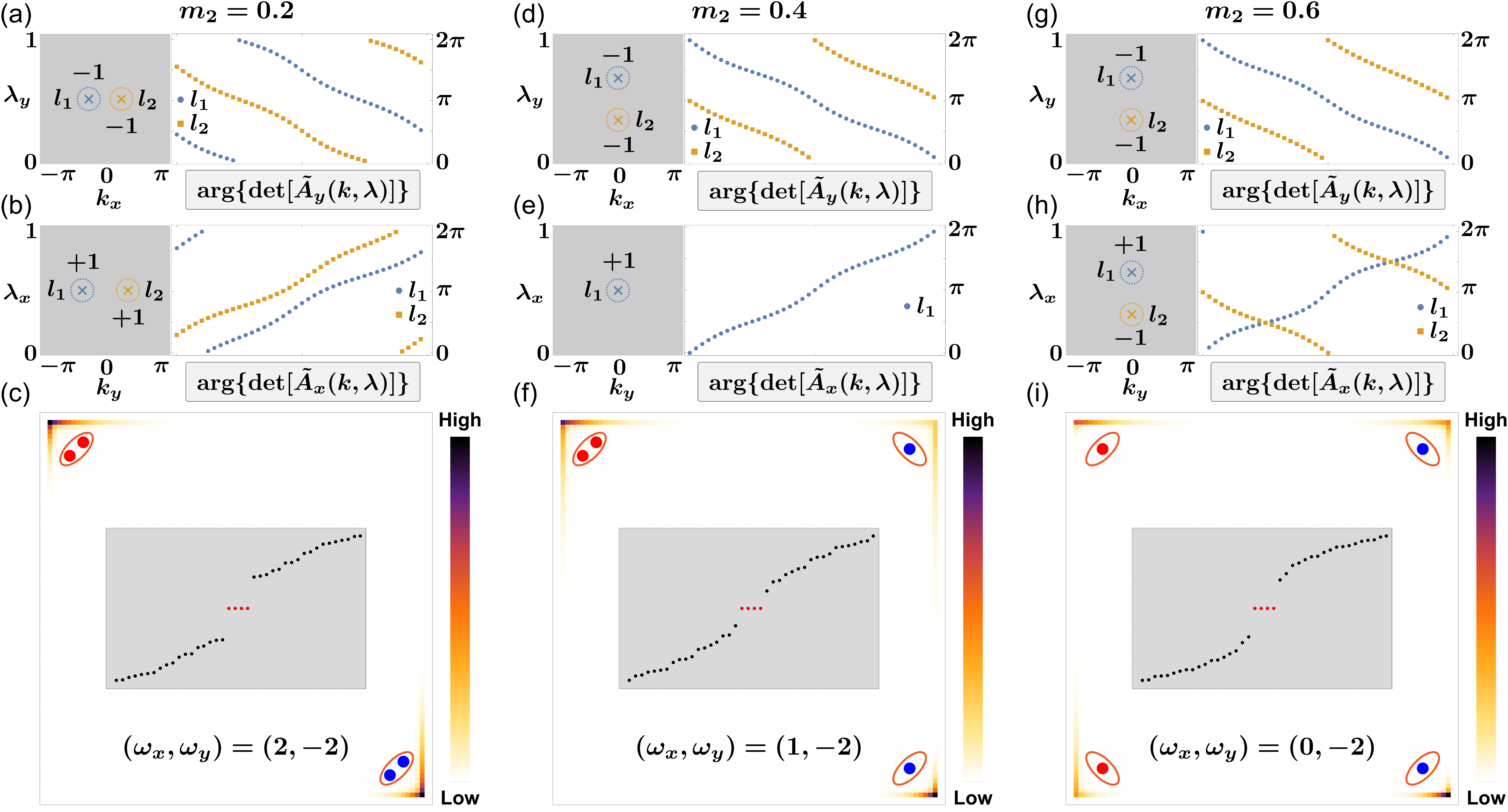}
    \caption{Zero-energy crossings and higher-order topological invariants in three distinct phases of the BDI model. The pair of higher-order invariants, $(\omega_x,\omega_y)$, is determined by topological charges of zero-energy crossings. In (a-b), (d)-(e) and (g)-(h), the crosses mark the relative positions of zero-energy crossings, and the numbers adjacent to each cross indicate their respective topological charges, which are obtained from winding numbers of the $D$ matrix along loops, $l_1$ and $l_2$, enclosing each crossing. (c), (f), (i) The distributions of Majorana corner modes. The solid blue and red circles represent Majorana zero modes with chiralities $+1$ and $-1$, respectively. Insets showcase energy spectra for an $80 \times 80$ open-boundary system. In these plots, $m_1=0.3t$ and $\delta m=0.1t$. }\label{fig9}
\end{figure*}

To establish the relationship between higher-order invariants and zero-energy crossings, we consider a large loop, $\mathcal C_3$, along the boundary of the parameter space. This loop consists of four line segments, $l_1$ to $l_4$, as depicted in Fig. \ref{fig8}(b). The winding number of the $D$ matrix along this giant loop is given by
\begin{align}
    \Omega & = \frac{1}{2\pi i}\sum_{i=1}^4\int_{l_i}dl \nabla_l \ln\det[D_a(k,\lambda)] \nonumber \\
    & = \sum_{(k_F,\lambda_F)} n_a(k_F,\lambda_F),
\end{align} 
where the summation in the last equality runs over all zero-energy crossings in the parameter space. The line integrals along $l_1$ and $l_3$ exactly cancel each other. We are thus left with integrals along $l_2$ and $l_4$, which essentially represent the winding number $\tilde\omega(1)$ and $\tilde\omega(0)$ as defined in Eq.(\ref{eq:winding_BDI}), up to a sign difference. Hence we have
\begin{equation}
    \Omega = \tilde\omega_a(0) - \tilde\omega_a(1).
\end{equation}
Given that $\tilde\omega_a(1)=0$, we now arrive at the relation between higher-order topological invariants and zero-energy crossings, which reads
\begin{equation}
    \omega_a = \sum_{(k_F,\lambda_F)} n_a(k_F,\lambda_F).\label{eq:topo_inv_BDI_1}
\end{equation}
It follows from Eq.(\ref{eq:topo_inv_BDI_1}) that the sign of the topological charge for each crossing matters, and that all the zero-energy crossings together determine the overall higher-order topological invariants.

\subsection{An example}

As a demonstration, we consider a 2D Hamiltonian in the BDI class, which is based on the model described in Eq.(\ref{eq:ham_DIII}). The Hamiltonian takes the form
\begin{equation}
    H_{\text{BDI}}(\bm k)  = H_{\text{DIII}}(\bm k) - m_2 \sigma_y.\label{eq:BDI_ham}
\end{equation}
Due to the additional $\sigma_y$ term, this Hamiltonian does not preserve time-reversal symmetry of the DIII-class Hamiltonian, which is $T=-i s_y\mathcal K$. Instead, the model exhibits a different time-reversal symmetry with $T = \sigma_x\mathcal K$. Since the latter time-reversal operator satisfies $T^2=1$, the Hamiltonian falls into BDI classification. The chiral symmetry operator is now represented as $S = \tau_x\sigma_x$. Similar to the DIII-class model, Hamiltonian in Eq.(\ref{eq:BDI_ham}) preserves inversion symmetry when $m_1$ and $\delta m$ terms are absent.

In the case where $m_1=m_2=0$, the system could possibly host a pair of zero-energy crossings on high-symmetry lines, depending on the relative strength of $m$ and $\delta m$. We already know that $m_1$ term doesn't couple the two crossings but instead separates them and drives them away from the high-symmetry lines. The $m_2$ term doesn't couple them either, albeit for a different reason. Due to the redefined chiral symmetry, the two crossings within the pair now carry the same topological charge, implying they cannot annihilate with each other. At the onset of $m_2$, the pair moves along the high-symmetry line in opposite directions. In this situation, turning on $m_1$ cannot immediately move each crossing off the line because of the time-reversal symmetry. However, for sufficiently large $m_1$, the two separated crossings on the high-symmetry line may meet again. If we continue to increase $m_1$, the pair would move off the line. Therefore, the positions of the two crossings vary according to the relative strength of $m_1$ and $m_2$. When the $m_1$ term is dominant, one might anticipate that the pair of zero-energy crossings occur off the high-symmetry lines at $(\pm k_F,\lambda_F)$. On the contrary, if the $m_2$ term is large compared to $m_1$, they may both reside on the high-symmetry lines, but at distinct points.

To illustrate these arguments, we present three representative cases with varying $m_2$ in Fig. \ref{fig9}. It's evident that the number and positions of zero-energy crossings, along with their topological charges, differ in each case, leading to distinct topological invariants. For smaller $m_2$, two crossings appear off the high-symmetry lines and have identical topological charges, as shown in Figs. \ref{fig9}(a)-(c). Following Eq.(\ref{eq:topo_inv_BDI_1}), we determine the topological invariants to be $(\omega_x,\omega_y)=(2,-2)$. As a result, four Majorana zero modes are found distributed across two opposite corners. At each corner, the two zero modes share the same chirality, as indicated by colors of the solid circles in Fig. \ref{fig9}(c). These topological invariants are further corroborated using Eq.(\ref{eq:chirality}), by noting that $N_{x-}^L=N_{y+}^L=0$ and $N_{x+}^L=N_{y-}^L=2$.

As $m_2$ increases, the two crossings are driven towards the high-symmetry line $K=0$ where they will meet with each other. Subsequently, the pair is pushed in opposite directions along the line. One of the crossings gradually approaches $\lambda=0$, and annihilates with its counterpart at negative $\lambda$. As depicted in Figs. \ref{fig9}(d) and (e) for $m=0.4$, two crossings are located on the line $K=0$ in the $(k_x,\lambda_y)$ space, whereas only one crossing are left in the $(k_y,\lambda_x)$ space as the other one has annihilated with the crossing with $-\lambda_x$ when they meet at $\lambda_x=0$. In this case, the topological invariants are $(\omega_x,\omega_y) = (1,-2)$. There are still four Majorana zero modes at the corners, but their distribution varies from the previous case. As shown in Fig. \ref{fig9}(f), two Majorana zero modes with chirality $-1$ group in one corner, while the others with chirality $+1$ occupy two separate corners, i.e., $N_{x+}^L=1$, $N_{x-}^L = N_{y+}^L = 0$ and $N_{y-}^L = 2$.

With a further increase in $m_2$, the two crossings in the $(k_x,\lambda_y)$ space remain. In the $(k_y,\lambda_x)$ space, however, another crossing emerges on the high-symmetry line carrying an opposite charge. We then have $(\omega_x,\omega_y)=(0,-2)$. Four Majorana modes are still present, but their locations differ from the previous cases. Now, each corner hosts one Majorana mode, with $N_{x+}^L=1$, $N_{x-}^L =1$, $N_{y+}^L = 0$ and $N_{y-}^L = 2$.

In all three cases, at least one of the two invariants is nonzero, indicating the nontrivial higher-order topology. It is possible that two systems with identical topological invariants can have distinct distributions, or even different numbers of Majorana corner modes. Nonetheless, the presence of nontrivial higher-order topology in BDI class is assured whenever at least one of the invariants assumes a nonzero value.

Like in the case of DIII class, the topological indicators are applicable regardless of whether the inversion symmetry is present. As we mentioned earlier in the DIII case, inversion symmetry guarantees that Majorana corner modes in the nontrivial higher-order phase ($\nu_x=\nu_y=1$) remain stable unless the bulk gap closes. It seems that in the BDI class, this argument may fail in certain cases. For example, we can start from the phase with Majorana modes at two corners, each of which hosts two zero modes with the same chirality, as in Fig. \ref{fig9}(c). It is possible that the two corners exchange one of the two modes while preserving inversion-symmetry as well as a finite bulk gap. Due to the difference in the chirality, the resulting two Majorana modes at the same corners annihilate each other. This inconsistency arises because the pair of $\mathbb Z$ invariants, $(\omega_x,\omega_y)$, characterize the so-called extrinsic classification \cite{geier2018}, where two phases are deemed to be topologically inequivalent if they cannot be smoothly transformed into each other while preserving both bulk and boundary gaps. In contrast, the intrinsic classification only requires the preservation of bulk gaps during the transformation, according to which the inversion-symmetry-protected BDI system is $\mathbb Z_2$ classified \cite{khalaf2018}. Considering that $\omega_x$ mod 2 = $\omega_y$ mod 2 in the presence of inversion symmetry, we may take $\omega_x(\omega_y)$ mod 2 to be the $\mathbb Z_2$ invariant characterizing the intrinsic classification.

\section{Discussions and conclusions}\label{sec:V}

In summary, we have demonstrated that zero-energy crossings, occurring when boundary conditions vary, can act as reliable indicators for higher-order topology in DIII and BDI classes. These crossings provide a direct and intuitive understanding of how bulk properties influence the emergence of Majorana zero modes at the corners. With variations in a bulk parameter, topological phase transitions may occur when either the number or the topological charges of these crossings change, even if the bulk gap remains open. The close relationship between zero-energy crossings and the bulk Hamiltonian directly illustrates the bulk-boundary correspondence in higher-order topological superconductors, irrespective of crystalline symmetries. The fact that zero-energy crossings could change when either the bulk or edge gap closes highlights a striking contrast to conventional bulk-boundary correspondence, where topological phase transitions typically occur only when the bulk gap closes.

In the DIII class, the $\mathbb Z_2$ charge for crossings on high-symmetry lines is obtained from the root's multiplicity by solving Eq.(\ref{eq:detA}) in the case of linear band crossings. For other types of crossings, this argument still holds. Let's first consider a pair of band crossings at $(K,\lambda_F)$, which means only $m_{a,1}$ in Eq.(\ref{eq:M_a}) equals zero at the crossing, and assume $m_{a,1} \propto (\lambda-\lambda_F)^r$ near the crossing. The exponent $r$ indicates the type of crossings. Apparently, only if $r$ is odd will $m_{a,1}$ change its sign after crossing $\lambda_F$. Hence, $\eta_a=1$ for odd $r$ and $\eta_a=0$ for even $r$. For instance, the quadratic band crossing corresponds to $r=2$, in which case $\eta_a$=0, suggesting that the Pfaffian of the $D$ matrix doesn't experience a $\pi$-phase jump across $\lambda_F$. Given that $\det[D_a(K,\lambda)]\propto (\lambda-\lambda_F)^{2r}$ near the crossing, we immediately obtain the root's multiplicity as $p_a=2r$. The parity of $p_a/2$ is exactly equal to that of $\eta_a$. For a generic case, $\eta_a$ is given by the number of $m_{a,j}$ that changes sign after crossing $\lambda_F$, which equals the number of crossing pairs with odd $r$. The root's multiplicity is expressed as $p_a = 2\rho_r r$, where $\rho_r$ represents the number of crossing pairs with exponent $r$. One can verify that $p_a/2 \text{ mod } 2 = \eta_a \text{ mod } 2$. Therefore, the $\mathbb Z_2$ charge of zero-energy crossings on the high-symmetry lines can be determined by the root's multiplicity, independent of the crossing type. For a crossing away from the high-symmetry lines, it is necessary to evaluate its topological charge defined in Eq.(\ref{eq:top_inv_DIII_A}), which carries an opposite charge to its time-reversal partner. In this case, the $\mathbb Z_2$ charge defined for the time-reversal pair is nonzero when the topological charge of either crossing within the pair is odd. An interesting question arises about whether one may determine the $\mathbb Z_2$ charge based solely on the root's multiplicity, akin to the crossings on high-symmetry lines. For example, consider two crossings off high-symmetry lines at $(\pm k_F,\lambda_F)$, each with multiplicity $p_a$. If we move these crossings towards the high-symmetry lines, which doesn't affect the $\mathbb Z_2$ charge, the multiplicity at the meeting point $(K,\lambda_F)$ on the high-symmetry lines becomes $2p_a$ unless the crossing type changes in this process. This implies that under these conditions, the $\mathbb Z_2$ charge could be inferred from the root's multiplicity. However, it's challenging to conclusively prove that the crossing type remains the same, which means we don't know whether $p_a$ remains constant before the crossing pair arrive at the high-symmetry lines, though it seems unlikely to change in most cases. In the model we studied in Sec. \ref{sec:III}D, $p_a$ for crossings off the high-symmetry lines doesn't change while the crossings move. In the BDI class, the distinction between zero-energy crossings on and off high-symmetry lines becomes irrelevant, and the root's multiplicity seems also to lose its significance. Here, the higher-order invariants rely on the cumulative topological charges of all crossings.

In this paper, we've characterized higher-order topology in 2D systems through pairs of topological invariants: $(\nu_x,\nu_y)$ for the DIII class, as defined in Eq.(\ref{eq:top_inv_DIII}), and $(\omega_x,\omega_y)$ for the BDI class, introduced in Eq.(\ref{eq:topo_inv_BDI_1}). A system enters nontrivial higher-order phases when at least one of these invariants becomes nonzero. However, the reverse isn't necessarily true. Nontrivial higher-order topology can occur even if both topological invariants are zero. For example, in the DIII class, if Majorana Kramers pairs are present at all four corners, the invariants may still be zero, i.e., $(\nu_x,\nu_y)=0$. This happens because these invariants represent the first-order topology of a quasi-1D cylindrical system, where each end corresponds to two corners in a square sample. When each corner supports a Majorana pair in the DIII class, their combined effect results in trivial invariants, due to the $\mathbb Z_2$ classification. A similar scenario occurs in the BDI class, where trivial invariants can arise if the two corners at each end host Majorana zero modes with opposite chiralities. This can also be viewed from the perspective of edge topology by noting that these invariants actually indicate the topological differences between opposite edges, as we have established in our previous study of D-class systems \cite{wang2023c}. However, in higher-order topological phases, topological differences may only exist between adjacent edges, rendering these invariants inadequate. For a comprehensive bulk-boundary correspondence in such cases, other general twisted boundary conditions, involving phase modulations of the boundary Hamiltonian, might be necessary \cite{niu1985,qi2006,song2020,zhang2022a,jahin2024,lin2024}. Our research underscores that boundary condition modulation can reveal a direct connection between bulk and corners. Future explorations within this framework may encompass a wider array of higher-order phases.

In systems with three dimensions (3D), higher-order topology can be second-order or third-order. To characterize these higher-order phases, we may explore the 3D parameter space and investigate zero-energy crossings within it. For second-order phases, we can consider variations in boundary conditions along one direction, with the parameter space spanned by two $k$ parameters and one $\lambda$ parameter, e.g., $(k_x,k_y,\lambda_z)$. Thus, three topological invariants are needed instead of two, as in the 2D case. If the quasi-2D boundary-modulated Hamiltonian, such as $H(k_x,k_y,\lambda_z=0)$, exhibits nontrivial first-order topology, a second-order topological phase featuring chiral or helical Majorana modes on the hinges will emerge. We also expect zero-energy crossings to occur in the $(k_x,k_y,\lambda_z)$ space. For third-order topology, variations in boundary conditions along two directions need to be considered, and the parameter space is characterized by one $k$ parameter and two $\lambda$ parameters, such as $(k_x,\lambda_y,\lambda_z)$. If the quasi-1D boundary-modulated Hamiltonian $H(k_x,\lambda_y=0,\lambda_z=0)$ is topologically nontrivial in the first-order sense, a nontrivial third-order phase with Majorana corner modes could arise. In this scenario, the zero-energy crossings are expected to extend into a line (ring) in the $\lambda_y\lambda_z$ plane for certain $k_x$. This is due to first-order topology of $H(k_x,\lambda_y=\pm 1,\lambda_z=\pm 1)$ being trivial, necessitating topological phase transitions as both $\lambda_y$ and $\lambda_z$ approach zero. Therefore, unlike in the second-order phase, where zero-energy crossings exist as nodal points, these crossings can form nodal lines in third-order phases. In the DIII and BDI classes, the third-order topology is determined by the topological charges of these nodal lines.

Finally, we make a comparison between the higher-order invariants introduced here and momentum-space topological invariants widely used to characterize first-order topological phases, such as the Chern number and the Fu-Kane invariant. Although the higher-order invariants proposed in this work can be obtained solely from the bulk Hamiltonian under the assumption of commensurate boundary terminations, they do not directly reflect the topology of the occupied state vector bundle defined over the bulk Brillouin zone. Unlike in conventional first-order phases, topological phase transitions in higher-order systems may occur while the bulk gap remains open, posing challenges in associating higher-order topological invariants with occupied bands over the bulk Brillouin zone. To address this issue, it is necessary to identify the qualitative (topological) changes in the bulk band for topological phase transitions occurring through the closing of boundary gaps, as demonstrated in Ref.\cite{jia2023,jia2024} for a class of minimal models. Additionally, it is worthwhile to examine other momentum-space topological indicators that reveal nontrivial topology in previously considered trivial phases and investigate their possible connections to higher-order topology \cite{wu2024c}.

\begin{acknowledgments}
    This work was supported by National Science Foundation of China (NSFC) under Grant No. 11704305, and the Innovation Program for Quantum Science and Technology (2021ZD0302400).
\end{acknowledgments}

\appendix* 
\section{The matrix $\tilde A_a$}\label{sec:appendix}

In this appendix, we will present the details in deriving the matrix $\tilde A_a(k,\lambda)$ introduced in Eq.(\ref{eq:tildeA}), whose determinant is shown to be equal to that of $A_a(k,\lambda)$ in Eq.(\ref{eq:Aa}).

Firstly, we note that the $D$ matrix of bulk Hamiltonian in momentum space is given by $D(\bm k) = u^\dagger H(\bm k)v$, where $U_S = (u\  v)$ is the unitary matrix that diagonalizes the chiral symmetry operator $S$. For the boundary-modulated Hamiltonian $H_a(k,\lambda)$ represented in the basis $\Psi_{a}(k) = \{\psi_{a,1}(k),...,\psi_{a,j}(k),...,\psi_{a,N_a}(k)\}$, with $\psi_{a,j}(k)$ including all internal degrees of freedom, denoted as $[c_a(k)]_{j\alpha}$, on the site $j$, the corresponding matrices $u$ and $v$ are replaced by $u_a = \mathbb I_{N_a\times N_a} \otimes u$ and $v_a = \mathbb I_{N_a\times N_a} \otimes v$, respectively. Here, $N_a$ represents the number of unit cells along the $a$ direction. As such, the $D$ matrix of boundary-modulated Hamiltonian is given by $D_a(k,\lambda) = u_a^\dagger H_a(k,\lambda) v_a$, and similarly for the boundary Hamiltonian $D_a^b(k) = u_a^\dagger B_a(k) v_a$. The entries of $A_a$ can be written as
\begin{align}
    [A_a(k,\lambda)]_{i\bar\alpha,j\bar\beta} &= \delta_{ij}\delta_{\bar\alpha \bar\beta}- (1-\lambda) \label{eqAppen:A}\\
    &\times [D_a^{-1}(k,1)]_{i\bar\alpha,l\bar\gamma} [D^b_a(k)]_{l\bar\gamma,j\bar\beta}, \nonumber
\end{align}
with Einstein summation assumed. In Eq.(\ref{eqAppen:A}) we distinguish $\bar \alpha$ and $\bar \beta$ from $\alpha$ and $\beta$ as the former indices run over only half the inner degrees of freedom. The two representations of bulk Hamiltonian, $D_a(k_{\bar a},1)$ and $D(\bm k)$, are related with each other through Fourier transformation, represented in matrix form as
\begin{equation}
    [D_a(k_{\bar a},1)]_{i\bar\alpha,j\bar\beta} =\sum_{k_a} [U_F^\dagger]_{i,k_a}[D(\bm k)]_{\bar \alpha,\bar\beta}[U_F]_{k_a,j}.\label{eqAppen:Da}
\end{equation} 
Here, the subscript of $k$ is explicitly shown for clarity, and the Fourier matrix $[U_F]_{k_a,j} = \frac{1}{\sqrt {N_a}} e^{-ijk_a}$. Since the boundary Hamiltonian operates only on boundary cells, it follows that the entries $[A_a(k,\lambda)]_{i\bar\alpha,j\bar\beta}=\delta_{ij}\delta_{\bar\alpha \bar\beta}$ if $j$ does not represent a boundary cell. Consequently, the determinant of $A_a$ can be calculated using only the block corresponding to boundary cells, denoted by $\bar A_a(k,\lambda)$, with
\begin{equation}
    \det[A_a(k,\lambda)] = \det[\bar A_a(k,\lambda)].\label{eqAppen:detBar}
\end{equation}
Combing Eqs.(\ref{eqAppen:A}) and (\ref{eqAppen:Da}), we arrive at
\begin{align}
    [\bar A_a & (k_{\bar a},\lambda_a)]_{i_b\bar\alpha,j_b\bar\beta}  = \delta_{i_b,j_b}\delta_{\bar\alpha,\bar\beta} - \frac{1}{2\pi}(1-\lambda_a)  \label{eqAppen:barA}\\
    & \times \int dk_a [F(k_a)]_{i_b,l_b} [D^{-1}(\bm k)]_{\bar\alpha,\bar\gamma} [\bar D^b_a(k_{\bar a})]_{l_b\bar\gamma,j_b\bar\beta},\nonumber
\end{align}
where $\bar D^b_a$ represents the block of $D^b_a$ corresponding to boundary cells labelled by $i_b$, $j_b$, $[F(k_a)]_{i_b,l_b}=e^{ik_a(i_b-l_b)}$, and the summation over $k_a$ is replaced by integration.

In certain cases, $\bar D^b_a(k)$ may not be a full-rank matrix. Its singular value decomposition is represented as
\begin{equation}
    [\bar D^b_a(k)]_{i_b\bar\alpha,j_b\bar\beta} = [\bar U_a^b(k)]_{i_b\bar\alpha,n}[\bar\Lambda_a^b(k)]_{n,n} [\bar V_a^{b\dagger}(k)]_{n,j_b\bar\beta},\label{eqAppen:Db}
\end{equation}
where $\bar\Lambda_a^b$ is a diagonal matrix with $r_a$ nonzero entries, with $r_a$ representing the rank of $\bar D^b_a(k)$. In addition, $\bar\Lambda_a^b$ can be arranged such that $[\bar\Lambda_a^b]_{n,n}\neq 0$ for $n\le r_a$. Given that $\bar U_a^b$ and $\bar V_a^b$ are unitary matrices, it follows that
\begin{equation}
    \det[\bar A_a(k,\lambda)]=\det[\bar V_a^{b\dagger}(k) \bar A_a(k,\lambda) \bar V_a^b(k)].
\end{equation}
From Eqs.(\ref{eqAppen:barA}) and (\ref{eqAppen:Db}) we know $[\bar V_a^{b\dagger} \bar A_a \bar V_a^b]_{m,n}=\delta_{m,n}$ if $n>r_a$. Hence we only need to consider the entries where $m,n \le r_a$ while calculating its determinant. For convenience, we group the first $r_a$ columns of $\bar V_a^b$ into a new matrix, denoted as $\tilde V_a^b$. Define
\begin{align}
    \tilde A_a(k,\lambda) = \tilde V_a^{b\dagger}(k) \bar A_a(k,\lambda) \tilde V_a^b(k), \label{eqAppen:tildeA}
\end{align}
and thus
\begin{equation}
    \det[\bar A_a(k,\lambda)]=\det[\tilde A_a(k,\lambda)].\label{eqAppen:detTilde}
\end{equation}
Combining Eqs.(\ref{eqAppen:barA}) and (\ref{eqAppen:tildeA}) we arrive at Eq.(\ref{eq:tildeA}) in the main text, which shows the specific form of this $r_a\times r_a$ matrix $\tilde A_a(k,\lambda)$.

\bibliography{flc.bib}

\end{document}